\documentclass[prd,twocolumn,amsmath,aps,floats,amssymb, floatfix, superscriptaddress, nofootinbib,preprintnumbers]{revtex4-1}

\usepackage{graphicx}
\usepackage{wasysym}
\usepackage{amsfonts}
\usepackage{subfigure}
\usepackage{ulem}
\usepackage[usenames]{color}
\definecolor{DarkGreen}{rgb}{0.0,0.5,0.0}

\def\beq{\begin{equation}}
\def\eeq{\end{equation}}
\def\be{\begin{equation}}
\def\ee{\end{equation}}
\def\bea{\begin{eqnarray}}
\def\eea{\end{eqnarray}}

\def\mpl{M_{\rm Pl}}

\newcommand{\goodgap}{\hspace{\subfigtopskip}\hspace{\subfigbottomskip}}

\def\r{\rho}

\def\e{\epsilon}

\def\m{\mu}
\def\n{\nu}

\def\r{\rho}
\def\s{\sigma}

\def\d{\partial}

\begin{document}


\preprint{IPMU-10-0089}
\preprint{NSF-KITP-10-069}

\title{Optimizing future experimental probes of inflation}
\medskip\
\author{ Damien A. Easson}%
\email[Email:]{easson@asu.edu}
\affiliation{ Department of Physics  \& School of Earth and Space Exploration  \& Beyond Center,
Arizona State University, Tempe, AZ 85287-1504, USA}
\affiliation{Institute for the Physics and Mathematics 
of the Universe, University of Tokyo,
5-1-5 Kashiwanoha, Kashiwa, Chiba 277-8568, Japan}
\affiliation{Kavli Institute for Theoretical Physics, University of California, Santa Barbara, CA 93106-4030, USA}
\author{Brian A.\ Powell} 
\email[Email:]{brian.powell007@gmail.com}
\affiliation{Institute for Defense Analyses, Alexandria, VA 22311, USA}

\begin{abstract}
The discovery of many novel realizations of the inflationary universe paradigm has led to a degeneracy problem: many
different inflationary Lagrangians generate the same perturbation spectra.
Resolving this problem requires the future
discovery of additional observables, beyond the scalar adiabatic and tensor two-point functions on CMB scales.
One important source of degeneracy arises in models where the density perturbation is generated by a non-inflationary degree of freedom, for example, through
curvatons or modulated reheating.  
We consider the curvaton scenario as representative of this class, and analyze the 
degeneracy with single field, canonical inflation that results if the curvaton goes undetected by future
observations. We perform Monte Carlo potential reconstructions in the absence of distinguishing observables, such as non-Gaussiantities
or isocurvature modes.  The resulting degeneracy is considerable and the improved measurements of spectral parameters from future
probes like CMBPol, offer little to better the situation.
Given a degeneracy-breaking observation, the observables must still be inverted to obtain the inflationary potential, with different observations resulting in reconstructions of varying quality.
We find that a future detection of isocurvature modes or a precision measurement of the tensor spectral index will enable
the most successful reconstructions in the presence of curvatons.  
\end{abstract}
\maketitle

\section{Introduction}
Observational cosmology has determined that the universe is homogenous and isotropic on scales larger than 100 Mpc and
that the universe is expanding according to Hubble's law. Measurements of the cosmic microwave background
radiation (CMB) reveal that this homogeneity and isotropy 
existed even at the time of recombination to an extreme accuracy at the order of $10^{-5}$ on all scales up to the present horizon. The modern standard model of cosmology attributes these observed properties of the 
universe to an early accelerating period, known as inflation~\cite{Guth:1980zm,Linde:1981mu,Albrecht:1982wi}. 

The inflaton, the field that dominates the energy density of the universe during inflation, traditionally assumes a dual role: it drives the accelerated expansion
and generates the primordial fluctuations that seed large scale structure.
Precision measurements of the CMB and large scale structure
surveys (LSS) have been instrumental in constraining the form of the inflationary
Lagrangian.  This process of utilizing cosmological observations to limit the form of
the inflationary Lagrangian is called {\it reconstruction}, and relies on the assumption
that there exists a unique mapping between the set of observables and the free parameters
of the Lagrangian.  Such a mapping exists in the framework of single field, canonical
inflation under the slow roll approximation, and the program of reconstruction has been
extensively developed \cite{Hodges:1990bf,Copeland:1993jj,Copeland:1993zn,Lidsey:1995np}.  
Of course, this process assumes that the inflaton is responsible for generating the perturbations; if this assumption is invalid then 
any process that seeks to `invert' a subset of observables to
obtain the underlying free parameters of the Lagrangian will reveal a space of
Lagrangians that is observationally degenerate~\footnote{We refer here to the
degeneracies in the locally reconstructed inflaton potential; degeneracies of a
different type arise globally as discussed in \cite{Linde:2005yw}.}: many distinct theories will provide the same
subset of observables \cite{Easson:2010zy}.  

It is possible to liberate the inflaton from its role in generating the primordial
perturbations.  In curvaton scenarios \cite{Linde:1996gt,Lyth:2001nq,Moroi:2001ct}, this role is assumed by an additional light degree
of freedom decoupled from the inflationary dynamics; in modulated reheating,
perturbations are generated by fluctuations in the decay rate of the inflaton
\cite{Dvali:2003em,Kofman:2003nx}.  In these
theories, the central assumption of traditional reconstruction--that the inflaton generates the
primordial spectra--is violated.  Without knowledge of how the spectra were generated,
whether by the inflaton or by some other means, a unique inversion of observables is
clearly impossible. 
The key to resolving this impasse is to recognize that while degeneracies might exist
within a subset of observables, it might be possible to utilize additional observables as
a means of distinguishing between different models.  While this is the obvious solution,
it is not clear which observables are relevant to which scenarios, or how
effectively such observations enable reconstruction.  It is now possible to state the {\it degeneracy problem}:
what observations must be made to distinguish between different theories, and if these observations
are not made how large is the resulting degeneracy?  The subsequent issue, one that we spend
considerable time investigating in this work, is the {\it inversion problem}: given a
distinguishing observation, how well can the set of observables be inverted to obtain the underlying
inflationary potential, $V(\phi)$?  

In this paper, we examine these questions within the context of the curvaton scenario as a prototype of the case in which the perturbation spectra
are not generated by the inflaton.  We do not consider modulated reheating, as it is phenomenologically similar to the
curvaton.
Our analysis employs Monte
Carlo reconstruction, in which the inflationary model space is stochastically sampled and models
satisfying specific observational constraints can be identified and extracted for further analysis.  We perform reconstructions for 
single field canonical inflation (serving as a reference model) and the curvaton scenario.
We first perform the reconstruction in the absence of any distinguishing observables to determine the size of the degeneracy, and then systematically investigate
the effects that an observation of isocurvature perturbations, non-Gaussianities, and a precision measurement of the tensor
spectral index each has on reducing the
degeneracy and enabling reconstruction.

This analysis is far from exhaustive; there are scores of inflationary scenarios beyond
single field inflation and curvatons that could conceivably contribute to the degeneracy
problem.  For example, the altered dynamical degrees of freedom of non-canonical inflation \cite{ArmendarizPicon:1999rj}
also contribute to the degeneracy problem \cite{Powell:2008bi}, and we investigate Dirac-Born-Infeld (DBI) inflation \cite{Silverstein:2003hf,Alishahiha:2004eh} in detail as an
example in \cite{debp10b}.  While other sources of degeneracy exist, we
consider the curvaton and DBI inflation to be two well-motivated, well-studied and
phenomenologically rich scenarios that are representative of two strong sources of
degeneracy: non-inflaton produced primordial spectra and non-canonical inflationary dynamics, respectively.
In the absence of a fully comprehensive analysis, the results presented here combined with 
\cite{debp10b} represent a `best-case' scenario for reconstruction, and should be viewed as a
first step towards understanding how large the degeneracy might be, and which
cosmological observables are needed to mitigate the problem. Despite the significant challenges we
identify in this study, it is our goal to determine which feasible cosmological observations will help us to best understand the fundamental
physics behind the inflationary universe paradigm. The information gathered in this analysis is crucial to the planning and
optimization of future cosmological experiments.
\section{Potential Reconstruction in Canonical Inflation}
\label{sec:recon}
The program of potential reconstruction was initiated in
1990 \cite{Hodges:1990bf}, and subsequently expanded and refined 
\cite{Copeland:1993jj,Copeland:1993zn,Lidsey:1995np}.  The reconstruction program initially targeted inflation driven by a
single, canonically normalized scalar field, $\phi$, minimally coupled to gravity,
\be
\frac{\mathcal L}{\sqrt{-g}} = \frac{\mpl^2}{2} R -  \frac{1}{2} g^{\m\n} \d_\m \phi \d_\n \phi - V(\phi)
\,.
\ee
The analysis provides a framework for mapping
power spectrum observables to the coefficients of a Taylor expanded inflaton potential.  The method begins with the
Hamilton-Jacobi equation,
\begin{equation}
\label{hje}
\frac{V(\phi)}{M_{\rm Pl}^2} = 3H(\phi)^2 - 2M_{\rm Pl}^2 [H'(\phi)]^2,
\end{equation}
where $H(\phi) \equiv \dot a/a$ is the Hubble parameter, $a(t)$ is the scale factor and $\prime \equiv d/d\phi$.  Next, the Hubble parameter is Taylor expanded
about some field value, $\phi_0$,
\begin{equation}
\label{series}
H(\phi) = \sum_{n = 0}^{\infty} \frac{1}{n!}\frac{d^nH}{d\phi^n}(\phi - \phi_0)^n.
\end{equation}
At zeroth-order, $H(\phi)$ is a constant, corresponding to the de Sitter solution.  The
first, second, and higher order terms are \cite{Liddle:1994dx}
\begin{eqnarray}
\label{derivs}
H' &=& \frac{H}{M_{\rm Pl}}\sqrt{\frac{\e}{2}},\\
H'' &=& \frac{H\eta}{2M_{\rm Pl}^2},\nonumber \\
&\vdots& \nonumber \\
\frac{d^{(n+1)}H}{d\phi^{(n+1)}} &=& \left(\frac{1}{2\mpl^2}\frac{H}{H'}\right)^n H' \lambda_n,\nonumber\\
\end{eqnarray}
where the functions $\epsilon = 2 \mpl^2 (H'/H)^2$, $\eta = \lambda_1$, and $\lambda_n$ ($n\geq 2$)
parameterize deviations from pure de Sitter expansion.  These are commonly referred to as
slow roll parameters in the literature; however, they are defined here without any assumption of
slow roll and will henceforth be referred to as {\it flow parameters}.  
Combining Eqs. (\ref{derivs}), (\ref{series}), and
(\ref{hje}) and collecting like-powers of $(\phi-\phi_0)$ leads to the Taylor coefficients of the potential,
\begin{eqnarray}
\label{pot1}
V(\phi_0) &=& 3M_{\rm Pl}^2 H^2(\phi_0),\\
\label{pot2}
V'(\phi_0) &=& \frac{3M_{\rm Pl}}{\sqrt{2}}H^2(\phi_0)
\sqrt{\epsilon(\phi_0)}\left[1-\frac{1}{3}\eta(\phi_0)\right],\\
\label{pot3}
V''(\phi_0) &=& 3H^2(\phi_0)\left[\epsilon(\phi_0) + \eta(\phi_0)\right.  \nonumber \\
& & \left.-\frac{1}{3}\eta^2(\phi_0)-\frac{1}{3}\epsilon(\phi_0)\,(\lambda_2(\phi_0))\right],\\
&\vdots& \nonumber
\end{eqnarray}
We note that these expressions are exact to all orders in slow roll.  

Potential reconstruction reveals the fact that
deviations from pure de Sitter are manifested in the power spectrum of density
fluctuations as deviations from scale invariance.  The power spectrum of curvature
perturbations can be parameterized as a power law,
\begin{equation}
P_\Phi(k) = P_\Phi(k_0) \left(\frac{k}{k_0}\right)^{n_s -1},
\end{equation}
where $k$ is the comoving wavenumber of the fluctuation (with $k_0$ defined as the scale that
crosses the horizon when $\phi = \phi_0$) and $n_s$ is the spectral index.
Deviations from pure de Sitter also result in the generation of a large-scale gravitational wave, or
tensor, spectrum:
\begin{equation}
P_h(k) = P_h(k_0) \left(\frac{k}{k_0}\right)^{n_T}.
\end{equation}
To first-order in slow roll, the amplitude of the power spectra at horizon crossing ($k=aH$) are given by the
well-known expressions,
\begin{eqnarray}
\label{amp}
P_\Phi(k) &=&  \left.\frac{1}{8\pi^2 M_{\rm Pl}^2}\frac{H^2}{\epsilon}\right|_{k=aH},\\
P_h(k) &=& \left.\frac{2}{\pi^2}\frac{H^2}{M_{\rm Pl}^2}\right|_{k=aH}.
\end{eqnarray}
The scale dependence of the spectra are then determined by the time variation of $H(\phi)$
and $\epsilon(\phi)$:
\begin{eqnarray}
\label{ns}
n_s(k_0) - 1 &\equiv& \frac{d{\rm ln}P_\Phi(k)}{d{\rm ln}k} = 2\eta(\phi_0) -
4\epsilon(\phi_0),\nonumber\\
\\
\label{nt1}
n_T(k_0) &\equiv&\frac{d{\rm ln}P_{h}(k)}{d{\rm ln}k} = -2\epsilon(\phi_0).
\end{eqnarray}
At first order in slow roll, the flow parameters $\epsilon$ and $\eta$ enable one
to map the spectrum observables, Eqs. (\ref{amp}-\ref{nt1}), to the Taylor coefficients
of the inflaton potential, Eqs. (\ref{pot1}-\ref{pot3}).  For example, at lowest order, we have
\begin{eqnarray}
\label{rlo1}
V(\phi_0) &=& \frac{\pi^2}{2}M_{\rm Pl}^4 P_\Phi(k_0)r(k_0),\\
\label{rlo2}
V'(\phi_0) &=& \frac{3}{4\sqrt{2}}\frac{V(\phi_0)}{M_{\rm Pl}} \sqrt{r(k_0)},\\
\label{rlo3}
V''(\phi_0) &=& \frac{3}{2}\frac{V(\phi_0)}{M_{\rm Pl}^2} \left[n_s(k_0)-1 + \frac{3}{8}r(k_0)\right],
\end{eqnarray}
where $r = P_h(k_0)/P_\Phi(k_0)$ is the tensor/scalar ratio, and we have neglected all
higher-order terms in Eqs. (\ref{pot1} -\ref{pot3}).  While only 
first-order approximations, Eqs. (\ref{rlo1} -\ref{rlo3}) provide a wealth of
information. For example, given the current measurement of the amplitude of scalar perturbations,
$P_\Phi(k_0)$, a detection of tensors by ESA's Planck Surveyor \cite{planck} will determine
$V(\phi_0)$ and $V'(\phi_0)$ to within a percentage error of roughly $\Delta r/r$.  The prospect of reconstructing
$V(\phi_0)$ is especially exciting because it corresponds to the energy scale of inflation -- an
important clue to uncovering the identity of the inflaton. 

The reconstruction equations (\ref{rlo1} -\ref{rlo3}) reveal that inflationary
potentials can be grouped into  three distinct
classes based on their observable predictions (c.f. Figure 1).
The economy afforded by this classification -- that vast numbers of different inflationary
potentials organize themselves into a few observational families -- came to be known as the `zoology' of inflationary models \cite{Dodelson:1997hr,Kinney:2003uw}.  
\begin{figure}
\label{regions}
\centerline{\includegraphics[width=3.5in]{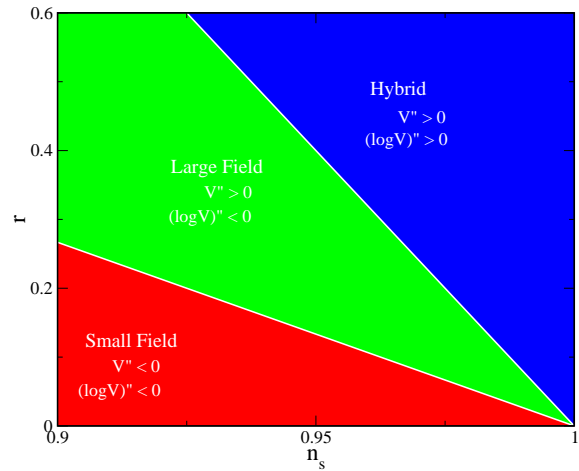}}
\caption{The `zoology' of inflation models.}
\end{figure}
Models labeled `hybrid' include potentials that evolve asymptotically to their minima,
requiring an auxiliary field to end inflation.  However, they are effectively single field models with
non-vanishing energy density at the minimum, and have the general form $V(\phi) \propto 1 + (\phi/\mu)^p$,
where $\mu$ is an energy scale and $p$ a positive integer. Hybrid models are characterized by the conditions $V''(\phi) > 0$ and $({\rm log}V(\phi))'' > 0$;  the simplest models of tree-level hybrid
inflation \cite{Linde:1993cn,Copeland:1994vg} belong to this class.  `Small field' and `large field' models are differentiated by
their initial field values.  Large field models are characterized by a field initially
displaced far from its minimum, with the general form $V(\phi) \propto (\phi/\mu)^p$, satisfying $V''(\phi) > 0$ and $({\rm log}V(\phi))'' < 0$; $m^2 \phi^2$ chaotic inflation \cite{Linde:1983gd} is one example.  Conversely, small
field models are characterized by a field initially close to the origin, with general form $V(\phi) \propto 1 - (\phi/\mu)^p$ (near the maximum), satisfying $V''(\phi) < 0$ and $({\rm log}V(\phi))'' < 0$; `new' inflation and
other models based on spontaneous symmetry breaking belong to this class.  While
phenomenologically distinct, these classifications are also useful for discerning more fundamental
aspects of inflation.  For example, large field models support chaotic initial conditions and eternal inflation, but are difficult to embed in a string theoretic framework;~\footnote{See, however, \cite{Silverstein:2008sg,McAllister:2008hb}} small field models with $\Delta \phi < M_{\rm Pl}$ are well-behaved effective potentials, but must have fine-tuned initial conditions; hybrid models are multifield scenarios.  

While future observations promise to reduce the size of errors in the $n_s$-$r$ plane, it
will still be possible to generate ensembles of potentials that provide good fits to the data.~\footnote{See,
for example, \cite{Kinney:2006qm,Lesgourgues:2007gp,Powell:2007gu}.}
However, with any luck, we will be able to determine which class of potential, of
the types discussed above, is ultimately responsible for driving inflation. 
This of course all assumes that the simplest implementation of single field, canonical,
minimally coupled inflation is true.  To what degree is our ability to reconstruct the
physics of inflation threatened by relaxing one or more of these assumptions?  In the
next section, we analyze the effects of a curvaton on the
reconstruction program. 
\section{Case Study: Reconstruction in the Presence of Curvatons}
\subsection{The Scenario and the Power Spectrum}
The curvaton mechanism \cite{Linde:1996gt,Lyth:2001nq,Moroi:2001ct} relaxes the assumption that the initial curvature perturbation
originated during inflation, and instead attributes it to the fluctuations of a
second field, $\sigma$,  known as the {\it curvaton}.  The curvaton is a weakly coupled scalar field that is relatively light during inflation, $m^2 \ll H^2$.  After inflation ends, the initially displaced curvaton rolls to its
minimum where it begins to oscillate during the post-inflationary radiation dominated phase.
These oscillations set up a small isocurvature perturbation that grows as the curvaton continues to oscillate.  After the curvaton decays, the
perturbation is converted to an adiabatic mode and structure begins to evolve according to the standard model.  The curvaton model is phenomenologically rich; depending on the energy density of the curvaton at the time of decay, there may be residual isocurvature modes or primordial non-Gaussianity large enough to be detected by future experiments.  
 
We first discuss the effect of the curvaton on the primordial power spectrum.  In general, both the curvaton and the
inflaton will contribute to the final curvature perturbation,
\begin{equation}
\label{pert1}
\Phi_f = -\frac{V}{M_{\rm Pl}^2V'}\delta \phi - \frac{3}{2}\frac{f(\sigma_*)}{M_{\rm Pl}}\delta \sigma_*,
\end{equation}
where $\Phi$ is the Bardeen potential (curvature perturbation in longitudinal gauge), $\delta \phi $ and $\delta
\sigma$ are the inflaton and curvaton vacuum fluctuations, and `*'
denotes the value at the end of inflation.  The function $f(\sigma)$ parameterizes the amount that the
curvaton contributes to the curvature perturbation.

In what follows, we will drop the subscripts `$f$' and `*' with the understanding that all
spectral observables are given during the radiation dominated phase following curvaton
decay.  The curvaton is effectively massless during inflation so that, $\delta \sigma = H/2\pi$, and the curvature perturbation Eq. (\ref{pert1}) becomes
\begin{equation}
\label{curvpert}
\Phi = \frac{1}{2\pi M_{\rm Pl}}\frac{H}{\sqrt{2\epsilon}} - \frac{3H}{4\pi M_{\rm Pl}}f(\sigma).
\end{equation}
Since the inflaton and curvaton perturbations
are uncorrelated, one can write the power spectrum \cite{Lyth:2002my,Langlois:2004nn,Moroi:2005kz},
\begin{equation}
\label{combspec}
P_\Phi = \left[1+\tilde{f}^2(\sigma)\epsilon\right]\frac{H^2}{8\pi^2\epsilon M^2_{\rm Pl}},
\end{equation}
where $\tilde{f} = 3f/\sqrt{2}$.  The spectral index follows,
\begin{equation}
\label{n_smod}
n_s - 1 \equiv \frac{d{\rm ln}P_\Phi(k)}{d{\rm ln}k} = -2\epsilon + \frac{2\eta - 2\epsilon}{1+\tilde{f}^2(\sigma)\epsilon}.
\end{equation}
Because the energy density of the curvaton is strongly subdominant during inflation it does not generate
gravitational waves, and the tensor spectrum is the same as that generated in single field inflation.  However, the tensor/scalar ratio is modified as a result of the new scalar amplitude Eq.(\ref{combspec}),
\begin{equation}
\label{rmod}
r = \frac{16\epsilon}{1+\tilde{f}^2(\sigma)\epsilon}.
\end{equation}
Note that by taking $\tilde{f} \rightarrow 0$ we recover the
usual inflaton-generated spectrum Eqs. (\ref{amp}) and (\ref{ns}), and so by tuning $\tilde{f}$ we can
control the contribution of the curvaton to the final perturbation.
The `pure curvaton' limit is achieved when $\tilde{f}^2(\sigma) \epsilon \rightarrow \infty$.

The addition of the curvaton alters the relationship between the observables
$r$ and $n_s$, allowing for a richer phenomenology.  By relegating
responsibility for generating a nearly scale invariant power spectrum to the curvaton,
constraints on the form of the inflaton potential are relaxed
\cite{Dimopoulos:2002kt,Moroi:2005np,Moroi:2005kz}.  However, it is also
apparent that the inclusion of the curvaton threatens a unique reconstruction
of the inflaton potential from observations.  The observables $r$ and $n_s$
pick up a dependence on $\tilde{f}(\sigma)$, and pass it along to the
potential coefficients. Solving for $\epsilon$ and $\eta$ in Eqs.
(\ref{n_smod})-(\ref{rmod}) and using these relations in Eqs. (\ref{pot1}-\ref{pot3}), we obtain new expressions for the potential coefficients,
\begin{eqnarray}
\label{crlo1}
V(\phi_0) &=& \frac{\pi^2}{2}M_{\rm Pl}^4 P_\Phi(k_0)r(k_0),\\
\label{crlo2}
V'(\phi_0) &=& \frac{3}{\sqrt{2}}\frac{V(\phi_0)}{M_{\rm
Pl}}\left(\frac{r(k_0)}{16-\tilde{f}^2(\sigma)r(k_0)}\right)^{1/2},\\ 
\label{crlo3}
V''(\phi_0) &=& \frac{V(\phi_0)}{M_{\rm
Pl}^2}\left(\frac{24}{16-\tilde{f}^2(\sigma)r(k_0)}\right)\nonumber \\
& & \times \left[n_s(k_0)-1 +
\frac{3}{8}r(k_0)\right].
\end{eqnarray}
Comparing Eqs. (\ref{rlo1}-\ref{rlo3}), we see that $V(\phi_0)$ is
unchanged by the inclusion of the curvaton because its energy density is subdominant to that of the inflaton.  
However, there is no longer a unique inversion from the observables $P_\Phi(k_0)$, $r$, and $n_s$ to the coefficients $V(\phi_0)$, $V'(\phi_0)$, and $V''(\phi_0)$, and the degree of resulting uncertainty depends on the possible values of $\tilde{f}(\sigma)$. 
In the limit $\tilde{f}^2(\sigma)\epsilon \gg 1$, the curvaton contribution dominates the overall curvature perturbation Eq. (\ref {curvpert}),  with the result that 
\begin{eqnarray}
r &\rightarrow& \frac{16}{\tilde{f}^2(\sigma)},\\
n_s -1&\rightarrow& -2\epsilon.
\end{eqnarray}
In this limit, the potential coefficients become
\begin{eqnarray}
\label{curvlimit}
V(\phi_0) &=& \frac{\pi^2}{2}M_{\rm Pl}^4 P_\Phi(k_0)r(k_0),\\                                                                       
V'(\phi_0) &=& \frac{3}{2}\frac{V(\phi_0)}{M_{\rm Pl}}\sqrt{1-n_s(k_0)},\\                                                           
\label{curvlimit3}
V''(\phi_0) &=& \frac{3}{2}\frac{V(\phi_0)}{M_{\rm Pl}^2}\left(\! \frac{dn_s/d{\rm ln}k(k_0)}{1-n_s(k_0)} + 1-n_s(k_0)\! \right)   
\end{eqnarray}
where $dn_s/d{\rm ln}k$ is the spectral index running.
A comparison between these expressions and Eqs.
(\ref{rlo1}-\ref{rlo3}) reveals that by tuning $\tilde{f}(\sigma)$ from `no curvaton contribution'
(single field) to `maximal contribution' (pure curvaton), we obtain very different relations between the inflaton potential and the observable parameters, with the possibility of the same potential giving rise to a wide range of  observables, depending on the value of $\tilde{f}(\sigma)$.  This is the degeneracy problem that threatens reconstruction in the presence of curvatons.  Equations (\ref{crlo1}-\ref{crlo3}) are derived from the lowest order
expressions for the observables $r$ and $n_s$; we neglected terms of order $\epsilon^2$, $\epsilon \eta$, etc. 
in obtaining Eqs. (\ref{n_smod}) and (\ref{rmod}).  These expressions thus serve merely to illustrate the emergence of possible degeneracies between
single field and curvaton models; we conduct a robust, higher-order reconstruction in {\S} \ref{MCs}.

In this analysis we allow a substantial range of values for $\tilde{f}(\sigma)$, including the limit in which the curvaton contribution is large. 
 This limit is attained if
the curvaton begins to oscillate early relative to the time that it decays; this way, the initially small curvaton fluctuations have time to grow.
Oscillations must then begin when the curvaton energy density is strongly subdominant: $H^2 \sim m^2 \gg \frac{m^2}{M_{\rm Pl}^2}\sigma^2$, and therefore the initial field
value must satisfy $\sigma \ll M_{\rm Pl}$.  
In this case, under the sudden decay approximation \cite{Lyth:2002my}, 
\begin{equation}
\label{limit}
\tilde{f}(\sigma) = r_\sigma\frac{4}{9}\frac{M_{\rm Pl}}{\sigma} ,\,\,\,\ (\sigma \ll M_{\rm Pl}),
\end{equation} 
where
\begin{equation}
\label{rsig}
r_\sigma = \frac{3\Omega_{\sigma,{\rm dec}}}{4 - \Omega_{\sigma,{\rm dec}}},
\end{equation}
and $\Omega_{\sigma,{\rm dec}}$ is the density parameter of the curvaton, $\Omega_\s \equiv \r_\s/\r$, at the time of
decay~\footnote{A more realistic gradual decay of the curvaton leads to similar results \cite{Malik:2006pm,Sasaki:2006kq}.}.  The contribution to the overall curvature perturbation is maximized when the curvaton comes to dominate the universe before decay, in which case $r_\sigma =\Omega_{\sigma,{\rm dec}} =1$. 

Note, that for $\sigma \sim M_{\rm Pl}$, the curvaton oscillates late; the contributions of the curvaton and inflaton to the final
curvature perturbation become comparable, and $\tilde{f}(\sigma)$ must be obtained numerically
\cite{Langlois:2004nn,Ferrer:2004nv}.  However, even for $\sigma \ll M_{\rm Pl}$, the curvaton can be subdominant
if $r_\sigma \ll 1$.  It is also possible that $\sigma \gg M_{\rm Pl}$, leading to a second bout of
curvaton-driven inflation after the inflaton energy density drops sufficiently.  This case is
essentially double inflation \cite{Silk:1986vc,Polarski:1992dq}, and we will not discuss it further.  

The curvaton scenario is phenomenologically rich, with the
ability to generate large non-Gaussianity and a significant isocurvature perturbation.  A detection of
either of these would serve to break the degeneracy with single field inflation.  The problem then
becomes one of inversion: if we are unable to constrain the function $\tilde{f}(\sigma)$, then potential
reconstruction will be degraded.  

\subsection{Isocurvature Modes}
\label{sec:iso}
As a multifield model, the curvaton scenario is capable of producing an isocurvature perturbation in addition to the
adiabatic perturbations discussed in the previous section.  A generic isocurvature perturbation between two separate
fluid components with energy densities $\rho_i$ and $\rho_j$ can be written,
\begin{equation}
\mathcal{S}_{ij} = 3(\zeta_i - \zeta_j),
\end{equation}
where we have introduced the curvature perturbation, $\zeta$, which is given during matter domination by $\zeta = -5\Phi/3$.
The curvaton can generate isocurvature perturbations in the decoupled fluids after inflation: cold dark matter (CDM),
baryons, and neutrinos \cite{Lyth:2002my},
\begin{eqnarray}
\mathcal{S}_{\rm CDM} &=& 3(\zeta_{\rm CDM} - \zeta_\gamma),\\ 
\mathcal{S}_{B} &=& 3(\zeta_{B} - \zeta_\gamma),\\ 
\mathcal{S}_{\nu} &=& 3(\zeta_{\nu} - \zeta_\gamma), 
\end{eqnarray}
where $\gamma$ denotes radiation.  
We define the power spectra
\begin{eqnarray}
\langle \zeta_{{\bf k}_1} \zeta_{{\bf k}_2}\rangle &\equiv& (2\pi)^3 \delta({\bf k}_1 + {\bf k}_2)P_\zeta(k_1),\\
\langle \mathcal{S}_{{\bf k}_1} \mathcal{S}_{{\bf k}_2}\rangle &\equiv& (2\pi)^3 \delta({\bf k}_1 + {\bf k}_2)P_\mathcal{S}(k_1),\\
\langle \zeta_{{\bf k}_1} \mathcal{S}_{{\bf k}_2}\rangle &\equiv& (2\pi)^3 \delta({\bf k}_1 + {\bf k}_2)P_{\zeta
\mathcal{S}}(k_1).
\end{eqnarray}
Most cosmological observations are sensitive to the relative amplitudes of the isocurvature and adiabatic spectra,
$|\mathcal{S}/\zeta|^2$, as well as the degree of correlation between the two modes.  The former quantity is conventionally measured in terms of the parameter,
\begin{equation}
\alpha \equiv \frac{\left|\frac{\mathcal{S}}{\zeta}\right|^2}{1+ \left|\frac{\mathcal{S}}{\zeta}\right|^2},
\end{equation}
and the latter~\footnote{Our sign convention is such that the temperature fluctuation 
$\Delta T/T = \zeta/5 - 2\mathcal{S}/5$ (take $\zeta = \mathcal{R} \rightarrow -\mathcal{R} = \tilde{\mathcal{R}}$ to compare with WMAP \cite{Komatsu:2008hk}).  Physically, a correlated perturbation ($\beta = 1$) corresponds to a suppression of the temperature power spectrum at the low multipoles.}
\begin{equation}
\beta \equiv {\rm cos}\Delta = \frac{P_{\zeta \mathcal{S}}(k_0)}{\sqrt{P_\zeta(k_0)P_\mathcal{S}(k_0)}}.
\end{equation}

A standard CDM isocurvature mode can be generated in either of two ways, depending on whether
the CDM was created before curvaton decay, or whether it was created as a direct product of this decay.  In the
former case, the CDM perturbation is generated by the inflaton, $\zeta_{\rm CDM} = \zeta_{\rm inf}$, and we obtain
\begin{equation}
\label{uncorr}
\left|\frac{\mathcal{S_{\rm CDM}}}{\zeta}\right|^2 = \frac{9}{1+\lambda^{-1}},
\end{equation}
where $\lambda = \tilde{f}^2(\sigma)\epsilon$, we have taken $\zeta_\gamma = \zeta$, and Eq.
(\ref{combspec}) has been used to relate the inflaton fluctuation to the overall curvature perturbation.  In the pure curvaton limit
($\tilde{f}^2(\sigma)\epsilon \rightarrow \infty$ and thus $\lambda \gg 1$), the isocurvature contribution is too large and has long
since been ruled out
\cite{Lyth:2002my}.  However, a sufficiently large inflaton contribution can reduce the amplitude of
$\mathcal{S}_{\rm CDM}$ and bring it into agreement with  current bounds;  this type of isocurvature perturbation is uncorrelated with
the adiabatic component ($\beta = 0$), with the constraint $\alpha_0 < 0.077$ (95\% CL) \cite{Komatsu:2010fb}. 
If future CMB missions observe a CDM isocurvature perturbation then single field
inflation will be ruled out and the degeneracy broken, and then a detection of tensors will
suffice to enable a successful inversion~\footnote{This assumes that the curvaton can be distinguished from
other mechanisms that might generate a CDM isocurvature mode, such as the axion.  This might require the simultaneous detection of non-Gaussianities consistent with the thermal scenario considered.}  -- we will have four functions ($V(\phi_0)$, $V'(\phi_0)$, $V''(\phi_0)$, $\tilde{f}(\sigma)$) in
terms of four observables ($P_\Phi(k_0)$, $r$, $n_s$, and $\alpha_0$).  The Planck surveyor is expected to obtain a
percentage error of around 4\% for CDM perturbations at 68\% CL \cite{Bucher:2000hy},  and we will analyze this case in detail in the next
section.  However, if the CDM is not generated before curvaton decay, then no such isocurvature mode will
be produced.

The other possibility is that the CDM is created as a direct product of curvaton decay.  
In this case \cite{Kawasaki:2008pa}
\begin{eqnarray}
\label{corr}
\left|\frac{\mathcal{S_{\rm CDM}}}{\zeta}\right|^2 &=& \frac{9(1 - r_\sigma)^2}{r_\sigma^2(1+\lambda^{-1})},\\
\beta &=& -\frac{1}{\sqrt{1+\lambda^{-1}}},
\end{eqnarray}
where $r_\sigma$ is as defined in
Eq. (\ref{rsig}).
In the pure curvaton limit ($\lambda^{-1} \rightarrow 0$), the isocurvature perturbation is anti-correlated ($\beta
= -1$), and its amplitude has been constrained in several analyses
\cite{Gordon:2002gv,Crotty:2003rz,Bucher:2004an,GarciaBellido:2004bb,Bean:2006qz,Kawasaki:2007mb},
most recently using WMAP7+BAO+SN
\cite{Komatsu:2010fb} giving $\alpha_{-1}  < 0.0037$ (95\% CL).  
From Eq. (\ref{rmod}), it is possible to relate the tensor/scalar ratio to the correlation angle,
\begin{equation}
\label{rb}
r = \frac{16\epsilon}{1 + \frac{\beta^2}{1-\beta^2}},
\end{equation}
and so anti-correlated CDM isocurvature modes correspond to a lack of tensors, $r=0$.
The loss of $r$ as an observable precludes a determination of the energy scale of inflation,
$V(\phi_0)$; however, the terms $V'/V$ and $V''/V$ are completely determined by $n_s$ and
$dn_s/d{\rm ln}k$ (c.f. Eqs. (\ref{curvlimit}-\ref{curvlimit3})).  We will investigate this case in the next section. 

While the un- and anti-correlated isocurvature perturbations discussed above are given the most
attention in the literature, it is evident from Eq. (\ref{corr}) that if the curvaton and inflaton
contributions to the curvature perturbation are both relevant, than it is possible to obtain a CDM
isocurvature mode with an arbitrary correlation~\footnote{See \cite{Valiviita:2009bp} for the most recent constraints on
arbitrarily correlated isocurvature perturbations.  Their sign convention on $\beta$ is opposite ours.}.  In this event, given measurements of the observables:
$r$, $n_s$, and $P_\Phi(k_0)$, a constraint on $\beta$ will determine $\tilde{f}(\sigma)$ and enable a
successful reconstruction, with Planck
expected to achieve $\Delta \beta \sim 0.04$ (65\% CL) \cite{Bucher:2000hy}.  We will investigate this
possibility in the next section. 

Before concluding this subsection, we mention that baryon isocurvature fluctuations appear identical to
CDM modes in the CMB, apart from a factor of $\Omega_b/\Omega_{CDM}$.  Therefore, the analyses that we
carry out in \S \ref{MCs} for the above cases involving CDM isocurvature perturbations apply to
baryonic modes as well.  

\subsection{Non-Gaussianity}
In addition to isocurvature modes, the multifield setting of the curvaton scenario has the potential to
generate large non-Gaussian temperature fluctuations in the CMB.
The non-Gaussianity of the fluctuations give rise to a nonzero bispectrum of the curvature
perturbation,
\begin{equation}
\label{bisp}
\langle \zeta^{\rm tot}_{{\bf k}_1}\zeta^{\rm tot}_{{\bf k}_2}\zeta^{\rm tot}_{{\bf k}_3}\rangle \equiv
(2\pi)^3 \delta({\bf k}_1+{\bf k}_2+{\bf k}_3)B^{\rm tot}_\zeta(k_1,k_2,k_3),
\end{equation}
where $\zeta^{\rm tot}$ denotes the total curvature perturbation in the matter dominated era, and
comprises both adiabatic and isocurvature components.  The bispectrum thus contains the four terms:
$\langle \zeta_{{\bf k}_1}\zeta_{{\bf k}_2}\zeta_{{\bf k}_3}\rangle$, $\langle \zeta_{{\bf
k}_1}\zeta_{{\bf k}_2}\mathcal{S}_{{\bf k}_3}\rangle$, $\langle \zeta_{{\bf k}_1}\mathcal{S}_{{\bf
k}_2}\mathcal{S}_{{\bf k}_3}\rangle$, and $\langle \mathcal{S}_{{\bf k}_1}\mathcal{S}_{{\bf
k}_2}\mathcal{S}_{{\bf k}_3}\rangle$ \cite{Kawasaki:2008sn,Langlois:2008vk,Kawasaki:2008pa}.    
In order to make contact with the parameter $f_{NL}$ that is constrained in contemporary analyses, we
define
\begin{equation}
B^{\rm tot}_\zeta (k_1,k_2,k_3) = \frac{6}{5}f_{NL}\left[P_\zeta(k_1)P_\zeta(k_2) + 2\,\, {\rm
perms}\right],
\end{equation}
with $\frac{6}{5}f_{NL} = \frac{6}{5}(f^{(3)}_{NL} + f^{\rm local}_{NL}) \approx \frac{6}{5}f^{\rm local}_{NL} = b_{\zeta \zeta \zeta} + \frac{1}{3}b_{\zeta \zeta \mathcal{S}} +
\frac{1}{9}b_{\zeta \mathcal{S} \mathcal{S}} + \frac{1}{27}b_{\mathcal{S} \mathcal{S} \mathcal{S}}$ where
$f^{(3)}_{NL}$ is related to the three-point function of the fields at horizon crossing, and $f^{\rm local}_{NL}$
arises from the nonlinear relation between
the curvature and field perturbations;  as we have indicated, the `local' form dominates the bispectrum in the
curvaton scenario.  The nonlinearity parameter, $f_{NL}$, characterizes the deviation of the fluctuation from pure
Gaussian, $\zeta = \zeta_g - \frac{3}{5}f_{NL}\left(\zeta^2_g - \langle\zeta^2_g\rangle\right)$, where $\zeta_g$ is
the Gaussian curvature perturbation.  Since it is local in real space, the Fourier modes satisfy ${\bf k}_1 \approx
{\bf k}_2 \gg {\bf k}_3$.
Using second-order perturbation theory
under the sudden decay approximation, the non-Gaussianity arising from the adiabatic component is \cite{Bartolo:2003jx,Lyth:2005du,Langlois:2010xc},
\begin{equation}
\label{ffNL}
f^{(\rm adi)}_{NL} = \frac{5}{6}b_{\zeta \zeta \zeta} = \frac{5}{6}\frac{1}{r_\sigma}\frac{\left(\frac{3}{2} - 2r_\sigma -
r_\sigma^2\right)}{\left(1 + \lambda^{-1}\right)^2}.
\end{equation}
The non-Gaussianity arising from the isocurvature component, $f^{(\rm iso)}_{NL} =
\frac{5}{6}\frac{1}{27}b_{\mathcal{S} \mathcal{S} \mathcal{S}}$, and that of the cross-correlated components
$\propto b_{\zeta \zeta \mathcal{S}}$ and $b_{\zeta \mathcal{S} \mathcal{S}}$, depends on the thermal
history of the universe: for the different CDM isocurvature modes considered in
the previous subsection we have $b^{\mathcal{S} \mathcal{S} \mathcal{S}}_{NL} = -3b^{\mathcal{S}
\mathcal{S} \zeta}_{NL} =9b^{\mathcal{S} \zeta \zeta}_{NL} = -27b^{\zeta \zeta \zeta}_{NL}$ for the
uncorrelated mode and $b^{\mathcal{S} \mathcal{S} \mathcal{S}}_{NL} \sim \frac{3}{r_\sigma}b^{\mathcal{S}              
\mathcal{S} \zeta}_{NL} \sim \frac{9}{r^2_\sigma}b^{\mathcal{S} \zeta \zeta}_{NL} \sim
\frac{27}{r^3_\sigma}b^{\zeta \zeta \zeta}_{NL}$ for the anti-correlated mode. 
Evidently, a sufficiently low curvaton energy density at the time of decay ($r_\sigma \ll 1$) will
result in large and potentially observable non-Gaussianities, and it is therefore possible for the
non-Gaussianity produced by the isocurvature mode to dominate the signal.  However, in the pure
curvaton limit it is in general difficult to generate large non-Gaussianities without simultaneously
violating the bounds on the isocurvature amplitudes.  For example, in the pure curvaton limit, the
bound on the anti-correlated mode ($\alpha_{-1}  < 0.0037$) corresponds to $r_\sigma \gtrsim 0.98$ and
a bound $f^{\rm local}_{NL} \in [-1.25,0.43]$.  This lies below the expected limits of future missions:  
the Planck surveyor is expected  to reach a 68\% limit of $\Delta f^{\rm local}_{NL} = 5$
\cite{Komatsu:2001rj} while large scale structure surveys should yield $\Delta f^{\rm local}_{NL} \sim 1
- 5$ \cite{Carbone:2010sb,Desjacques:2010nn} at 68\% CL. Simultaneously observable non-Gaussianities and isocurvature contributions
are, however, possible for certain neutrino \cite{Gordon:2003hw} and baryonic
modes in the case of mixed curvaton and inflaton perturbations \cite{Moroi:2008nn}. 
However, as discussed in the last subsection, a detection of isocurvature modes independently imposes
a constraint on the function $\tilde{f}(\sigma)$, and measurements of
non-Gaussianities are not needed for reconstruction. 
A detection of $f_{NL}$ is, however, especially advantageous in the event that no isocurvature perturbations
are produced in the scenario because it will break degeneracies and might help with inversion. 

In the absence of isocurvature production, the only nonzero contribution to the non-Gaussianity is
from $f^{\rm (adi)}_{NL}$, which is a function of both $r_\sigma$ and $\tilde{f}(\sigma)$.  In order
to constrain $\tilde{f}(\sigma)$, an additional, complementary observation is needed.
In addition to the bispectrum, the non-Gaussian fluctuations generate a trispectrum statistic
\cite{Hu:2001fa,Seery:2006vu,Byrnes:2006vq},
\begin{equation}
\langle \zeta_{{\bf k}_1}\zeta_{{\bf k}_2}\zeta_{{\bf k}_3}\zeta_{{\bf k}_4}\rangle \equiv
(2\pi)^3 \delta\left(\sum_{i=1}^4{\bf k}_i\right)T_\zeta(k_1,k_2,k_3,k_4),
\end{equation}
where, in contrast to Eq. (\ref{bisp}), we include only the adiabatic perturbations since we are
considering the case in which isocurvature modes are absent.~\footnote{See \cite{Kawakami:2009iu} for the full
trispectrum including isocurvature contributions.}  The trispectrum can be written in terms of the
estimators $\tau_{NL}$ and $g_{NL}$ \cite{Okamoto:2002ik,Kogo:2006kh},
\begin{eqnarray} 
T_\zeta(k_1,k_2,k_3,k_4) &=& \tau_{NL}(P_\zeta(k_1)P_\zeta(k_{13})P_\zeta(k_4) + 11\,\,{\rm
perms})\nonumber\\
&+& \frac{54}{25}g_{NL}(P_\zeta(k_2)P_\zeta(k_3)P_\zeta(k_4) + 3\,\,{\rm perms}).\nonumber \\
\end{eqnarray}
The function $g_{NL}$ contains higher-derivative terms and is subdominant relative to $\tau_{NL}$ \cite{Sasaki:2006kq},
and we will not consider it further.  
The estimator $\tau_{NL}$ can be related to the physical parameters of the model using the $\delta N$
formalism \cite{Lyth:2005fi,Byrnes:2006vq,Ichikawa:2008iq},
\begin{equation}
\tau_{NL} = \frac{36}{25}f^2_{NL}(1+\lambda^{-1})^2.
\end{equation} 
Happily, the trispectrum does not introduce additional free parameters, and a measurement of $f^{\rm
local}_{NL}$ and $\tau^{\rm local}_{NL}$ will constrain the functions $\tilde{f}(\sigma)$ and $r_\sigma$.  
The 95\% limit on the bispectrum from WMAP7  is $-10 < f^{\rm local}_{NL} < 74$ \cite{Komatsu:2010fb}, while the
trispectrum is much less well-constrained: $|\tau_{NL}| \lesssim 10^8$ from COBE \cite{Boubekeur:2005fj}.  Future
full-sky galaxy surveys might achieve an accuracy of $\Delta f^{\rm local}_{NL} \sim 1-5$ (68\% CL)
\cite{Carbone:2010sb,Desjacques:2010nn} and
Planck is expected to achieve $\Delta \tau^{\rm local}_{NL} \sim 560$ (95\% CL) \cite{Kogo:2006kh}.
We will study the effect of such detections on reconstruction in \S \ref{MCs}. 

In general, it is possible for the amplitude of non-Gaussianity, $f^{\rm local}_{NL}$, to vary with
scale, $f_{NL} \sim k^{n_{NG}}$, where the spectral index is defined
\begin{equation}
n_{NG} = \frac{df_{NL}}{d{\rm ln}k} \,.
\end{equation}
Scale dependent local non-Gaussianities \cite{Byrnes:2008zy} can arise in the curvaton
scenario if the curvaton potential has interaction terms \cite{Byrnes:2010xd} or if the final density
perturbation is a comparable mixture of inflaton and curvaton fluctuations \cite{Byrnes:2009pe}.  Since we assume
the conventional quadratic curvaton potential, $V(\sigma) \propto m^2\sigma^2$, we consider the latter possibility for which
\begin{equation}
\label{nng}
n_{NG} = 4\left[1-\frac{1}{1+\lambda^{-1}}\right](2\epsilon - \eta + \eta_\sigma),
\end{equation}
where $\eta_\sigma \propto H_{\sigma \sigma}/H \ll \eta$, since $H^2 \gg \rho_\sigma$
during inflation.  The projected 1$\sigma$ error on $n_{NG}$ from future CMB missions is
\cite{Sefusatti:2009xu}
\begin{equation} 
\label{errornng}
\Delta n_{NG} \simeq 0.1\frac{50}{f^{\rm local}_{NL}}
\end{equation}
for a full-sky survey with Planck, and a factor of two smaller with CMBPol.  The prospects of measuring a scale dependence are exciting because
reconstruction may proceed without input from the amplitude $f^{\rm local}_{NL}$, since $n_{NG}$ by itself determines $\tilde{f}^2(\sigma)$.  We consider
such a detection in \S \ref{MCs}.

Lastly, we discuss the possibility that neither isocurvature modes nor non-Gaussianities are detected;
in this case a precision measurement of the tensor spectral index can be leveraged to assist
in the reconstruction effort.

\subsection{Tensor Index}
\label{curvgrav}
The energy density of the curvaton field during inflation is too small to generate a detectable tensor perturbation, and hence
the primordial gravitational wave spectrum is instead determined by the dynamics of the inflaton field, 
\begin{eqnarray}
P_h &=& \frac{2}{\pi^2}\frac{H^2}{M_{\rm Pl}^2},\\
\label{nt}
n_T &=& -2\epsilon.
\end{eqnarray}
However, since the curvaton does contribute to the overall density perturbation, the tensor/scalar ratio is given by Eq. (\ref{rmod}) and the single field consistency relation, $r = -8n_T$, is modified:
\begin{equation}
\label{modcons}
r = \frac{-16n_T}{2 - \tilde{f}^2(\sigma)n_T}.
\end{equation}

A detection of $r$ and $n_T$ can be used to break degeneracies with single field
inflation if their relationship is sufficiently different from the single field consistency relation
$r=-8n_T$ \cite{Song:2003ca,Smith:2006xf}.  However, even if a particular observation
confidently rules out single field inflation, it might not be clear what the alternative theory is since, in addition to curvatons, multifield
\cite{Wands:2002bn,Sasaki:1995aw} and
DBI inflation \cite{Silverstein:2003hf,Alishahiha:2004eh}, and even trans-Planckian effects
\cite{Hui:2001ce,Kaloper:2002uj} lead to an altered consistency relation.  
An accurate identification of the underlying theory might require additional corroborating
observations, like the amplitude and shape of non-Gaussianities or isocurvature modes.  
Our intent is not to enumerate all the different theories that predict the same relationship between
$r$ and $n_T$, and so we will not view a measurement of $n_T$ as a degeneracy-breaking observation.
Its status as such is mostly irrelevant for our purposes: as long as the measurements of $r$ and $n_T$
are consistent with the consistency relation in question to within experimental error, it can be
used to constrain $\tilde{f}(\sigma)$.

Reconstruction is possible because a determination of
$\tilde{f}(\sigma)$ enables the inversion -- four functions ($V(\phi_0)$, $V'(\phi_0)$, $V''(\phi_0)$, 
$\tilde{f}(\sigma)$) in terms of four observables ($P_\Phi(k_0)$, $r$, $n_s$, and $n_T$).  
In fact, from Eq. (\ref{modcons}) and Eqs. (\ref{crlo1}-\ref{crlo3}) we find that $V'/V
\propto \sqrt{\epsilon} \propto \sqrt{-n_T}$ for the curvaton. Meanwhile, the single field reconstruction $V'/V \propto \sqrt{\epsilon}$ can be written
either in terms of $r$ or $n_T$ by virtue of the single field consistency relation.   The relative error between the curvaton
and single field reconstructions of $V'/V$ is then
\begin{equation}
\label{nT_recon}
\frac{\Delta (V'/V)_{\rm curv}}{\Delta (V'/V)_{\rm single}} = 8\frac{\Delta n_T}{\Delta r},
\end{equation}
where the assumption is that $r$ is better constrained than $n_T$; otherwise $\epsilon$ would be most accurately determined by $n_T$ in both models, and the
reconstructions would be equivalent.
The 
other reconstruction diagnostic -- the zoology classification -- depends on the uncertainty in $\tilde{f}^2(\sigma)$:
\begin{equation}
\Delta \tilde{f}^2(\sigma) = \frac{1}{r}\frac{16}{1+\Delta r/r} + \frac{1}{n_T}\frac{2}{1-\Delta n_T/n_T}.
\end{equation}
We see that when $\Delta (V'/V)_{\rm curv} = \Delta(V'/V)_{\rm single}$, the zoology reduces to that of single field inflation.
An accurate detection of $n_T$
will, however, be difficult with current technology: the Planck satellite in combination with current ground-based experiments like
QUIET \cite{Samtleben:2008rb}, BICEP \cite{Yoon:2006jc}, and PolarBear \cite{polarbear}, might achieve a 1-$\sigma$ error of
$\Delta n_T \sim 0.1$ \cite{Zhao:2009rt}.  While future space-based platforms such as CMBPol
\cite{Bock:2006yf,Baumann:2008aq,Bock:2009xw} might reduce this error by a factor of $2$ in the absence of foregrounds \cite{Verde:2005ff}, a precision
measurement of $n_T$ will most likely require the direct detection of primordial gravitational waves.

The prospect of a direct detection of primordial gravitational waves on scales $\sim 0.1-1$ Hz is the focus of
several recent concept studies using space-based laser interferometers.  Two proposals that have seen much analytical attention are
the Big Bang Observer (BBO) \cite{bbo} and Japan's DECIGO project \cite{Seto:2001qf}.  The forecasts established in these
studies indicate that while BBO will be competitive with Planck ($\Delta n_T \sim 0.1$), upgrades
to BBO might achieve $\Delta n_T \sim 10^{-2}$, and DECIGO might reach $\Delta n_T \sim
10^{-3}$ \cite{Seto:2005qy,Kudoh:2005as} in the most optimistic cases.  These missions are still in the concept stage, and even if they become a reality,
data will not become available for at least another decade.  Even so, we include the possibility of a
tensor index measurement for completeness.
We now investigate the accuracy of the potential reconstruction that follows from a
detection of each of the observables discussed in \S \ref{sec:iso}-\ref{curvgrav}.   

\subsection{Monte Carlo Analysis}
\label{MCs}
The mapping from observables to potential parameters, for example  Eqs. (\ref{rlo1} - \ref{rlo3}) in the case
of slow roll inflation, is a lowest order result.  We wish to obtain a higher-order reconstruction; however,
while the task of inverting the observables ($r$, $n_s$, $dn_s/d{\rm ln}k$, $\cdots$) to obtain the flow parameters
($\epsilon$, $\eta$, $\lambda_2$, $\cdots$) in order to reconstruct the potential (via Eqs. (\ref{pot1} -
\ref{pot3})) is tractable at low order, it becomes prohibitively difficult
at higher order.
This difficulty motivated the development of Monte Carlo
reconstruction \cite{Easther:2002rw} in which one begins with the flow
parameters: $\epsilon$, $\eta$, $\lambda_2$, $\cdots$, $\lambda_M$, taken as initial
conditions, and for each randomly drawn set obtains the inflationary
evolution by solving the set of flow equations
\cite{Hoffman:2000ue,Kinney:2002qn},
\begin{eqnarray}
\label{Flow}
\frac{dH}{dN} &=& \epsilon H ,\nonumber \\
\frac{d\epsilon}{dN} &=& 2\epsilon(\eta - \epsilon), \nonumber \\
\frac{d\lambda_\ell}{dN} &=& [(\ell -1)\eta -\ell \epsilon]\lambda_\ell + \lambda_{\ell +1} ,\,\,\,\ \ell \in [1,M],\nonumber \\
\frac{d{\rm ln}k}{dN} &=& \epsilon - 1,
\end{eqnarray}
where the system is truncated by taking $\lambda_{M+1} = 0$.  Our time
variable, $dN = -Hdt$, is the number of efolds before the end of inflation.
In traditional applications
\cite{Hoffman:2000ue,Kinney:2002qn,Moroi:2005np,Kinney:2006qm,Powell:2007gu},
only solutions that yield sufficient inflation to solve the horizon and
flatness problems are retained, typically with $|\Delta N| \in [45,70]$.
However, in this study, we impose the relatively diminished restriction $|\Delta N| \geq 10$, as this corresponds to the time period during which length scales directly probed by cosmological experiments exit the horizon.  We work to $6^{th}$-order in slow roll, and we have verified that our conclusions are robust up to $10^{th}$-order.  
For each inflationary solution, observables and potential coefficients are separately determined.
There is no need for a difficult inversion; the flow parameters are stochastically sampled and
models corresponding to an observation of interest can then be selected out.  The result is a
consistent, high-order reconstruction that improves upon the analytic reconstruction given by Eqs. (\ref{rlo1}-\ref{rlo3}).  This approach, of course, generalizes to include the curvaton.            
In order to obtain a robust determination of possible degeneracies, it is                                
important to sample a wide array of different inflation models and compare                               
their potential reconstructions under the assumptions of both single field                               
inflation and the curvaton model.

We perform two flow analyses: one for single field inflation in which the inflaton is soley responsible for generating primordial perturbations, and the other for the curvaton scenario, in which the curvaton contributes to primordial perturbations.  
For the case of single field inflation, we draw the initial flow parameters uniformly from the ranges 
\begin{eqnarray}
\label{ics}
\epsilon_i &\in& [0,0.8],\nonumber \\
\lambda_{\ell,i} &\in& 10^{-\ell + 1}[-0.5,0.5],
\end{eqnarray}
and we take ${\rm ln}k_i = -8.047$ so that the largest scale corresponds to the quadrupole.  
We solve the flow equations and if $|\Delta N| \geq 10$, we calculate the potential parameters Eqs. (\ref{pot1}-\ref{pot3}), and  the observables \cite{Stewart:1993bc},
\begin{eqnarray}
\label{canobs}
r &=& 16\epsilon [1 - 2C(\eta - \epsilon)],\\
n_s - 1 &=& 2\eta - 4\epsilon - 2(1 + C)\epsilon^2 \nonumber \\
&&- \frac{1}{2}(3-5C)\epsilon \eta + \frac{1}{2}(3-C)\lambda_2,\\
\frac{d n_s}{d{\rm ln}k} &=& 4\epsilon(\eta - \epsilon) -{4\epsilon^2 - 12\epsilon \eta + 2\lambda_2},\\
n_T &=& -2\epsilon - (3 + C)\epsilon^2 + (1+C)\epsilon \eta,
\end{eqnarray}
at $k = 0.01 {\rm Mpc}^{-1}$.  Here $C = 4({\rm ln}2 + \gamma) - 5 \approx 0.0815$ and $\gamma$ is the
Euler-Mascheroni constant.  We have included the running of the spectral index in
order to obtain a higher-order reconstruction.

For the curvaton scenario, we draw the flow parameters from the ranges Eq. (\ref{ics}) and uniformly
sample $\tilde{f}(\sigma) \in [25,10,000]$.  The lower bound on
$\tilde{f}(\sigma)$ is chosen so that the limit Eq. (\ref{limit}) is accurate, which can be verified
by consulting the numerical solution for $\tilde{f}(\sigma)$ in \cite{Langlois:2004nn}, while the
upper bound is arbitrary.  We solve the
flow equations and if $|\Delta N| \geq 10$, we calculate the potential coefficients Eqs. (\ref{crlo1}-\ref{crlo3}) and the observables,
\begin{eqnarray}
\label{cr}
r &=& \frac{16\epsilon [1 - 2C(\eta - \epsilon)]}{1+\tilde{f}^2(\sigma)\epsilon},\\
\label{cn_s}
n_s - 1 &=& -2\epsilon + \frac{2\eta - 2\epsilon}{1+\tilde{f}^2(\sigma)\epsilon} - 2(1 + C)\epsilon^2
\nonumber \\
& &- \frac{1}{2}(3-5C)\epsilon \eta + \frac{1}{2}(3-C)\lambda_2,\\
\label{ca}
\frac{d n_s}{d{\rm ln}k} &=& 4\epsilon(\eta - \epsilon) -\frac{4\epsilon^2 - 12\epsilon \eta +
2\lambda_2}{1+\tilde{f}^2(\sigma)\epsilon} + \frac{4 \tilde{f}^2(\sigma)\epsilon(\eta -
\epsilon)^2}{(1+\tilde{f}^2(\sigma)\epsilon)^2},\nonumber \\
\\
n_T &=& -2\epsilon - (3 + C)\epsilon^2 + (1+C)\epsilon \eta,
\end{eqnarray}
at $k = 0.01 {\rm Mpc}^{-1}$.
We now present results for the different observational outcomes and collect our findings in Table
1.

\subsubsection{No Detection of Non-Gaussianity or Isocurvature Modes}
\label{base}
\begin{figure*}
\centering
$\begin{array}{ccc}
\subfigure[]{
\includegraphics[width=0.32 \textwidth,clip]{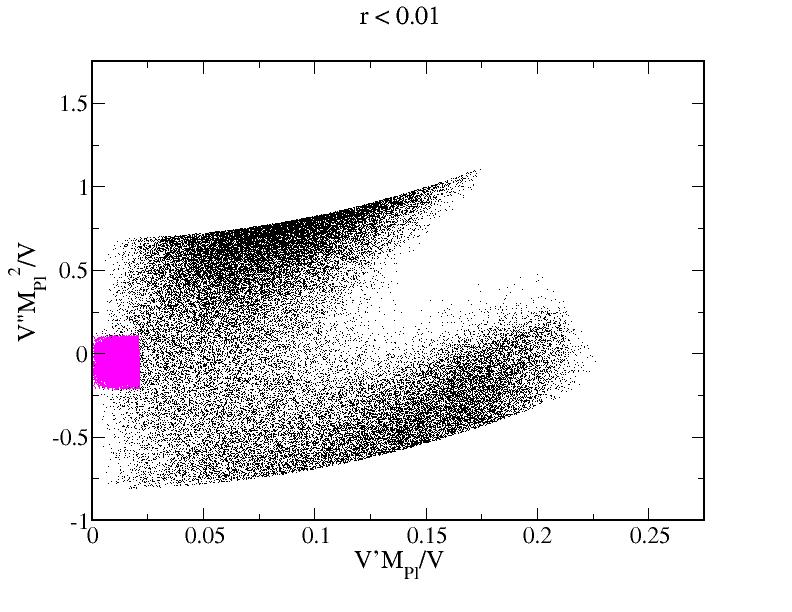}}
\subfigure[]{
\includegraphics[width=0.32 \textwidth,clip]{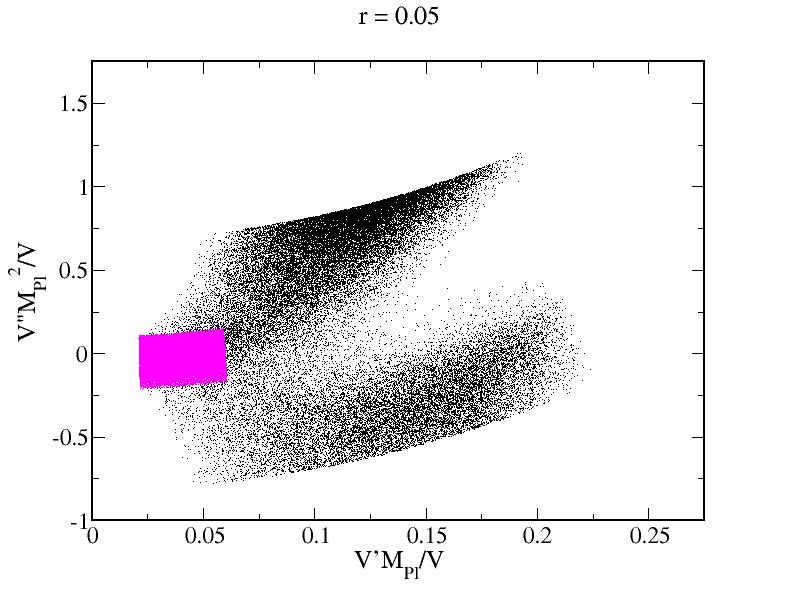}}
\subfigure[]{
\includegraphics[width=0.32 \textwidth,clip]{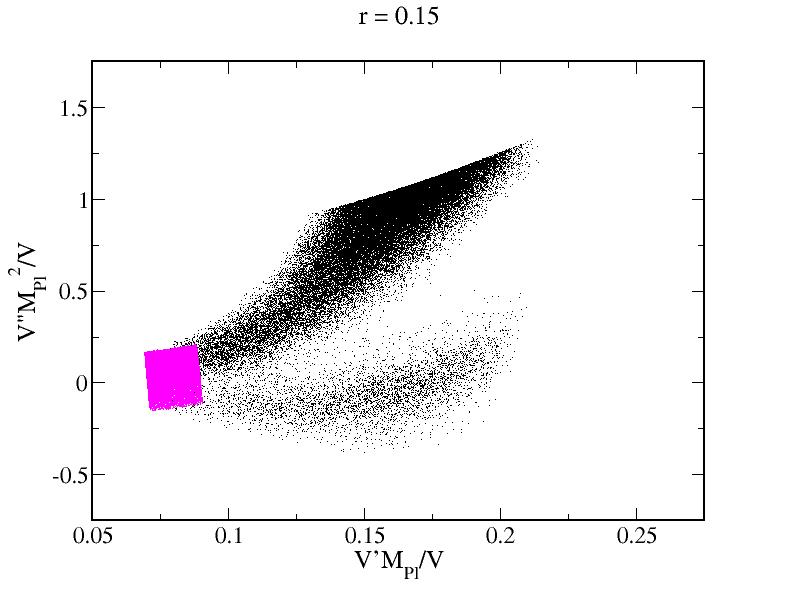}}
\end{array}$
$\begin{array}{ccc}
\subfigure[]{
\includegraphics[width=0.32 \textwidth,clip]{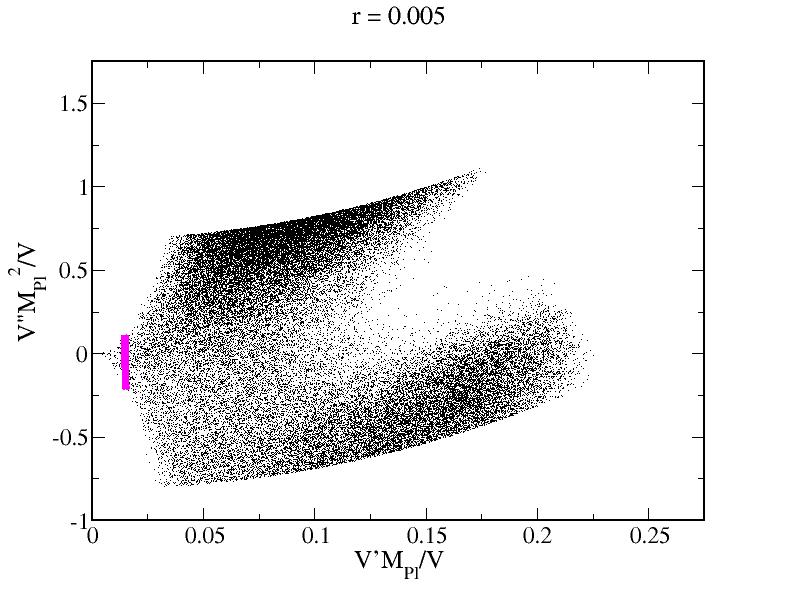}}
\subfigure[]{
\includegraphics[width=0.32 \textwidth,clip]{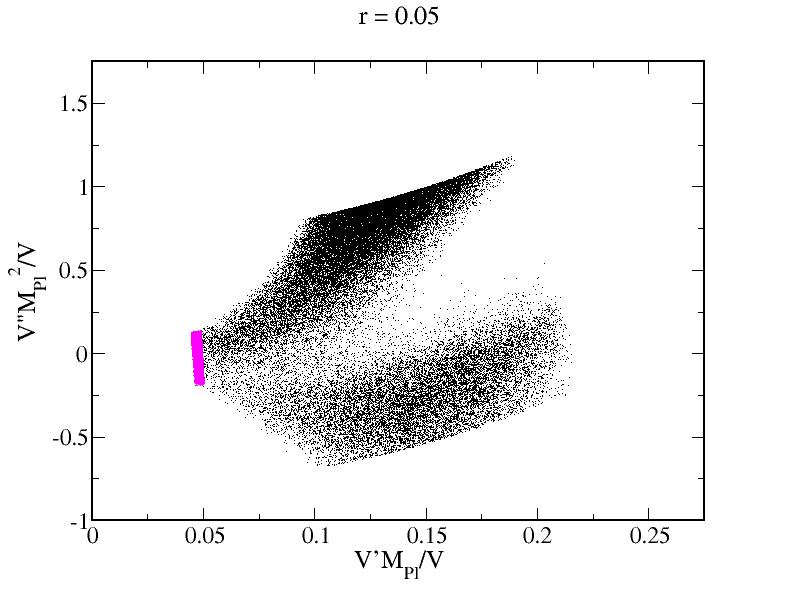}}
\subfigure[]{
\includegraphics[width=0.32 \textwidth,clip]{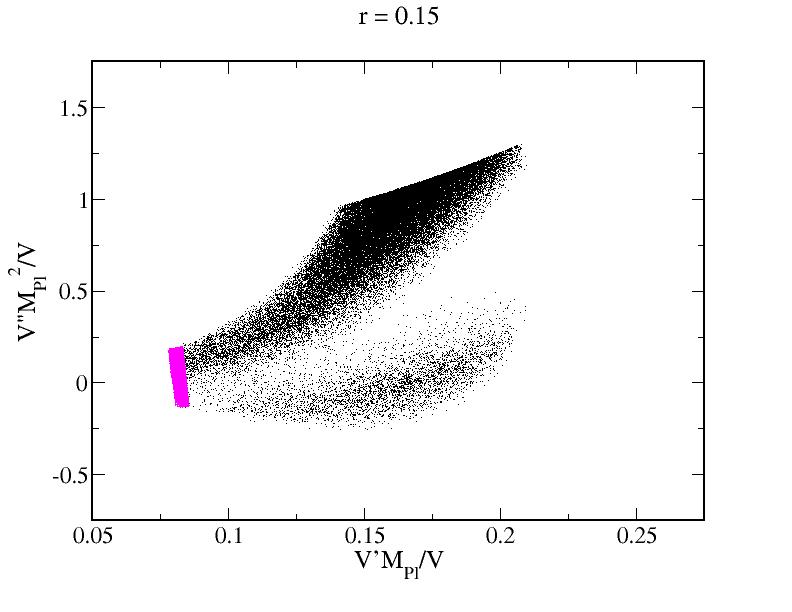}}
\end{array}$
\caption{Monte Carlo results of single field (magenta) and curvaton (black) reconstructions in the absence of
non-Gaussianities, isocurvature modes, or a precision measurement of $n_T$.  We
present results for three fiducial values of $r$: $r = 0.005$ (a),(d), $r = 0.05$ (b),(e), $r = 0.15$ (c),(f).  Top row presents results for Planck and bottom
row for CMBPol.}
\end{figure*}
We first make projections for reconstruction utilizing expectations
from both the Planck Surveyor and CMBPol for the case in which neither isocurvature perturbations nor non-Gaussianities are
observed.  We do not include the possibility of a precision measurement of $n_T$ at this time.  This can be
considered the worst-case scenario in which the degeneracy persists, and shall serve as a baseline against which to compare different observational
outcomes.  For Planck, we assume 68\% CL detections of $r$ ($r
\gtrsim 0.01$, $\Delta r \sim 0.03$) \cite{Colombo:2008ta}, $n_s$ ($\Delta
n_s \sim 0.0038$), and $dn_s/d{\rm ln}k$ ($\Delta dn_s/d{\rm ln}k \sim 0.005$)
\cite{Bond:2004rt}.  
For CMBPol, we assume 68\% CL detections of $r$ ($r \gtrsim 10^{-4}$, $\Delta r \sim r/10$), $n_s$ ($\Delta n_s \sim 0.0016$), and $dn_s/d{\rm ln}k$
($\Delta dn_s/d{\rm ln}k \sim 0.0036$) \cite{Baumann:2008aq}.
We will assume this base set of observations throughout the remainder of this analysis, unless
otherwise indicated.  The tensor
spectral index will not be adequately constrained with these CMB missions and the modified consistency
relation will not be useful for constraining curvaton models.  

We performed a $6^{th}$-order flow analysis on both single field inflation
and the curvaton scenario,                                     
collecting 50,000 models of each.  We plot our results in Figure $2$ for three 
fiducial values of $r$: $r=0.005$ (a,d), $r = 0.05$ (b,e) and $r = 0.15$ (c,f), with Planck on top
and CMBPol on bottom.  Since (a) lies below the Planck detection threshold, in this case Planck
can only impose the upper bound $r \lesssim 0.01$. 
The magenta points represent single field models and the black points
curvaton models.  Figure 2 is a robust, higher-order representation of
the degeneracies first mentioned in the context of Eqs. (\ref{crlo1}-\ref{crlo3}); the uncertainty in the potential parameters in
the curvaton scenario is a result of our inability to constrain
$\tilde{f}^2(\sigma)$.  
One can
view the curvaton as `contaminating' the single field result -- by allowing
for the presence of curvatons in the reconstruction, constraints are
degraded relative to single field inflation by a factor of around four for
$V'/V$ and a factor of five for $V''/V$.  
We assume fiducial values for the spectral index and running: $n_s = 0.97$ and $dn_s/d{\rm ln}k =0$, although there is little sensitivity to the values of $n_s$ and $dn_s/d{\rm ln}k$ chosen: changing $n_s$ acts to
shift the distributions slightly up and down, while changing $dn_s/d{\rm ln}k$ has no noticeable effect.~\footnote{This is because $V'(\phi_0)$ and $V''(\phi_0)$ depend only weakly on running ($V'''(\phi_0)$ depends more
strongly); it is therefore unlikely that even an accurate determination of $dn_s/d{\rm ln}k$, as might be possible
with future 21-cm line observations \cite{Barger:2008ii}, will be helpful.}

The inclusion of the curvaton does not affect the reconstruction of the height of the potential,
$V(\phi_0)$.  The higher-order reconstruction is consistent with the conclusion drawn analytically
at lowest-order.  The lack of a
tensor detection by Planck ($r \lesssim 0.01$) will therefore only impose an upper bound on $V(\phi_0)$, but does not
otherwise strongly affect the errors on $V'/V$ and $V''/V$, as seen by comparison of Figure 2 (a) with (b).  If Planck fails to constrain $r$, future
concepts like CMBPol offer the tantalizing possibility of detecting B-modes with $r \gtrsim 10^{-4}$.
In such a case ($r=0.005$), the single field reconstruction is notably improved (magenta models in Figure 2 (d)):  $V'/V$, which is directly
proportional to $\sqrt{r}$, improves by a factor of 5 over the case in which Planck fails to detect $r$.  However, surprisingly, such a detection has virtually
{\it no effect} on the curvaton reconstruction (black models in Figure 2 (d).)  This is generally true at the other fiducial $r$ values as seen by
comparison of Figures 2 (b) with (e) and (c) with (f), due to the poor constraints on $\tilde{f}^2(\sigma)$ in the degenerate
case; the percentage error on $V'/V$ for curvatons is
\begin{equation}
\frac{\Delta (V'/V)}{V'/V} \propto \left[2\left(\frac{\Delta r}{r}\right)^2 + \left(\frac{\Delta \tilde{f}^2(\sigma)}{\tilde{f}^2(\sigma)}\right)^2\right]^{1/2}.
\end{equation}
While $\Delta r/r \lesssim 1$ for CMBPol, the error on $\tilde{f}^2(\sigma)$ is large: $\Delta \tilde{f}^2(\sigma)/\tilde{f}^2(\sigma) \approx 4000$, and easily
overwhelms the constraints on $r$.  We thus reach an important conclusion: {\it in the
degenerate case, the sole benefit of a detection of $r$ will be the determination of the energy scale of
inflation; the constraints on $V'/V$ and $V''/V$ are largely insensitive to a detection or lack of
detection of $r$, even by more advanced probes like CMBPol.}\footnote{In this statement we are assuming that background pollution of the tensor
signal can be accurately subtracted. Detection of $r$ is, of course, significant in that it provides unprecedented evidence for the existence of gravitational waves.}  Nevertheless, in what follows we will continue to include $r$ as a base observable because it will become important
when we consider measurements of $n_T$ \S \ref{curvddgw}.

As a result of the curvaton degeneracy, a given observation is consistent with a greater variety of
potentials.  Recall that in single field inflation, for example,  models that predict $r < -8(n_s-1)/3$ satisfy
$V'' < 0$ and $({\rm log}V)'' < 0$ and are classified as small field potentials.  In light of the
effect of the degeneracy, it is now unclear whether regions in the observable
$n_s$-$r$ plane map uniquely to classes of functions of similar form.
The extent to
which an unresolved curvaton preserves this mapping and the resulting model classification
provides an alternative view of the degeneracy problem, complementary to the constraints found on
$V(\phi)$ in Figure 2.
\begin{figure*}[htp]
\centering
$\begin{array}{cc}
\subfigure[]{
\includegraphics[width=0.49 \textwidth,clip]{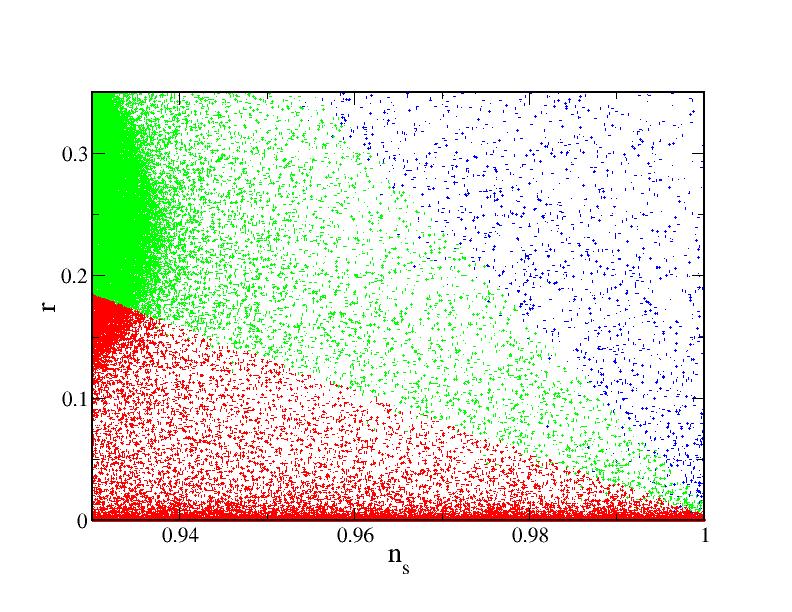}}
\subfigure[]{
\includegraphics[width=0.49 \textwidth,clip]{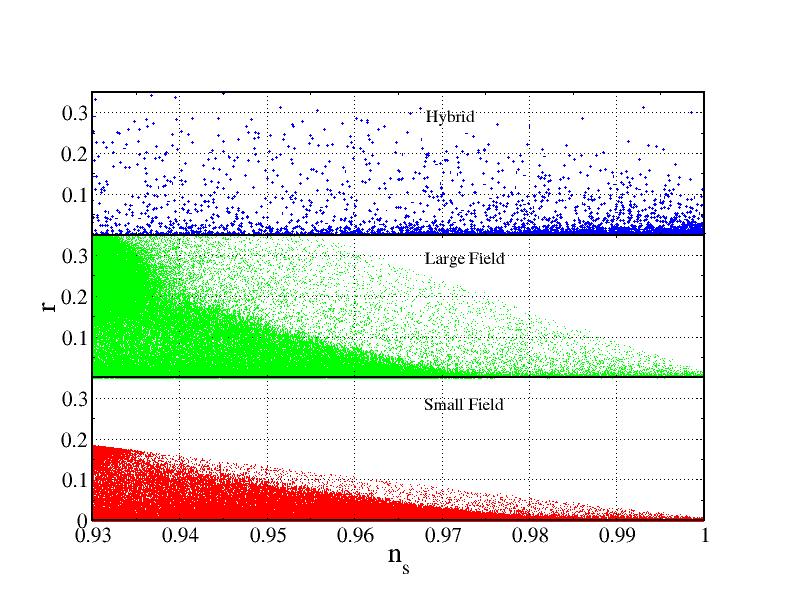}}
\end{array}$
\caption{(a) Zoology of single field models in the $n_s$-$r$ plane, obtained using the flow formalism (c.f. Figure 1).  Blue, green,
and red models represent hybrid, large field, and small field potentials, as defined in {\S \ref{sec:recon}}.
(b) Zoology of models in which a curvaton is present but remains undetected.  Because the classes overlap, we plot each model-type separately.  Small field models are
completely degenerate with large field models.  Both types are completely degenerate with hybrid models.}\end{figure*}
We generate a new zoology in the presence of curvatons by assigning each of the models in Figure 2
to a zoology class (small field, large field, hybrid) as per the discussion at the end of \S \ref{sec:recon}. 
We first apply this methodology to single field models, and reproduce in Figure 3 (a) the zoology introduced in
Figure 1.  The slight overlap in regions is a result of the higher-order reconstruction of our analysis.  In
Figure 3 (b), we
present the modified zoology that results when the curvaton is also present. In both cases we collect models
within the observable ranges set by the WMAP7 95\% CL: $r < 0.55$, $n_s \in [0.91,1.2]$, and $dn_s/d{\rm ln}k \in
[-0.09, 0.02]$ \cite{Komatsu:2010fb}.  The zoology is most clearly presented with each class plotted separately, so as to
avoid overlap.  We find that a substantial region of parameter space is degenerate in
classification.  In particular, {\it all observables compatible with large field models are also found to be
consistent with hybrid models}.  The
boundary between small and large field models lies at $V''(\phi_0) = 0$.  At lowest order, this occurs when
$\epsilon = -\eta$ (c.f. Eqs. (\ref{pot1}-\ref{pot3})), which gives a boundary in the observable plane:
\begin{equation}
\label{line1}
r = 12\tilde{f}^{-2} - 2\tau \pm 2\tilde{f}^{-2}\left[\left(6-\tilde{f}^2\tau\right)^2 + 16\tilde{f}^2\right]^{1/2},
\end{equation}
where $\tau = n_s -1 $.  The boundary evidently depends on the value of $\tilde{f}$: as $\tilde{f} \rightarrow
\infty$, this curve becomes more horizontal, approaching the line  $r = 0$.  While only a lowest order approximation,
this result gives some intuition as to why the large field models extend all the way to the line $r=0$ in
Figure 3 (b).  

The same approach applied to the boundary between hybrid and large field models, given by $({\rm
log}V(\phi))'' = 0$, and thus $\epsilon = \eta$, leads to the relation
\begin{equation}
\label{2}
r = \frac{16\tau}{\tilde{f}^2\tau - 2}.
\end{equation}
This curve also tends to the line $r=0$ as $\tilde{f} \rightarrow \infty$, and coincides with the single field
case in the limit $\tilde{f} \rightarrow 0$.  This agrees with our second finding,
that {\it all observations compatible with small field models are compatible with both large field and hybrid
models.}  Hybrid models populate the full observable parameter space, extending down through the large field
and small field regions.~\footnote{In \cite{Kobayashi:2009nv} an explicit hybrid potential
is studied that yields $r=0$, $n_s < 0$, typical small field observables, using
curvatons/modulated reheating.}  Only those hybrid models existing in the single field `hybrid' region can be
correctly classified in the presence of the curvaton, i.e. they must  satisfy $r > 8(1 -n_s)$. 
For perspective, we present the latest marginalized constraints on $n_s$ and $r$
from WMAP+BAO+$H_0$ \cite{Komatsu:2010fb} in Figure 4.  Only observables outside the yellow region, a minority of the high
CL                                                                 
parameter space, can be uniquely classified according to the inflationary zoology.
Note that,
as in single field models, a detection of observables with $r>  8(1 -n_s)/3$ would rule out small field inflation models.

It is particularly notable that large field curvaton models can accommodate a small tensor amplitude on observable scales.  
Recall that in single field inflation, the relation
\begin{equation}
\frac{d\phi}{dN} = M_{\rm Pl}\sqrt{2\epsilon},
\end{equation}
together with $r = 16\epsilon$ imposes a lower bound on the variation of the inflaton field during inflation,
\begin{equation}
\Delta \phi \geq M_{\rm Pl}\sqrt{\frac{r}{8}}\Delta N,
\end{equation}
where $\Delta N$ corresponds to observable scales.
A tensor amplitude satisfying $r \lesssim 0.01$ indicates that $\Delta \phi < M_{\rm Pl}$ while observable scales left the horizon.~\footnote{Of course, it is still possible that $\Delta \phi > \mpl$ over the full course of inflation, even when $r < 0.01$
\cite{Efstathiou:2005tq,Easther:2006qu}.}
In the presence of the curvaton the bound is
modified:
\begin{equation}
\Delta \phi = M_{\rm Pl}\sqrt{\frac{r}{8}}\left(\frac{16}{16-\tilde{f}^2(\sigma)r}\right)^{1/2}\Delta N.
\end{equation}
As $\tilde{f}^2(\sigma)r \rightarrow 16$, the field variation grows without bound.  From Eq. (\ref{rmod}), this 
correspond to the limit of {\it small} $r$: in the presence of curvatons, $r< 0.01$ no longer implies that $\Delta \phi < \mpl$
even across observable scales.~\footnote{Information about the field range may be recovered if $\tilde f(\sigma)$ can
be accurately constrained.}

\begin{figure}
\centerline{\includegraphics[width=2.5in]{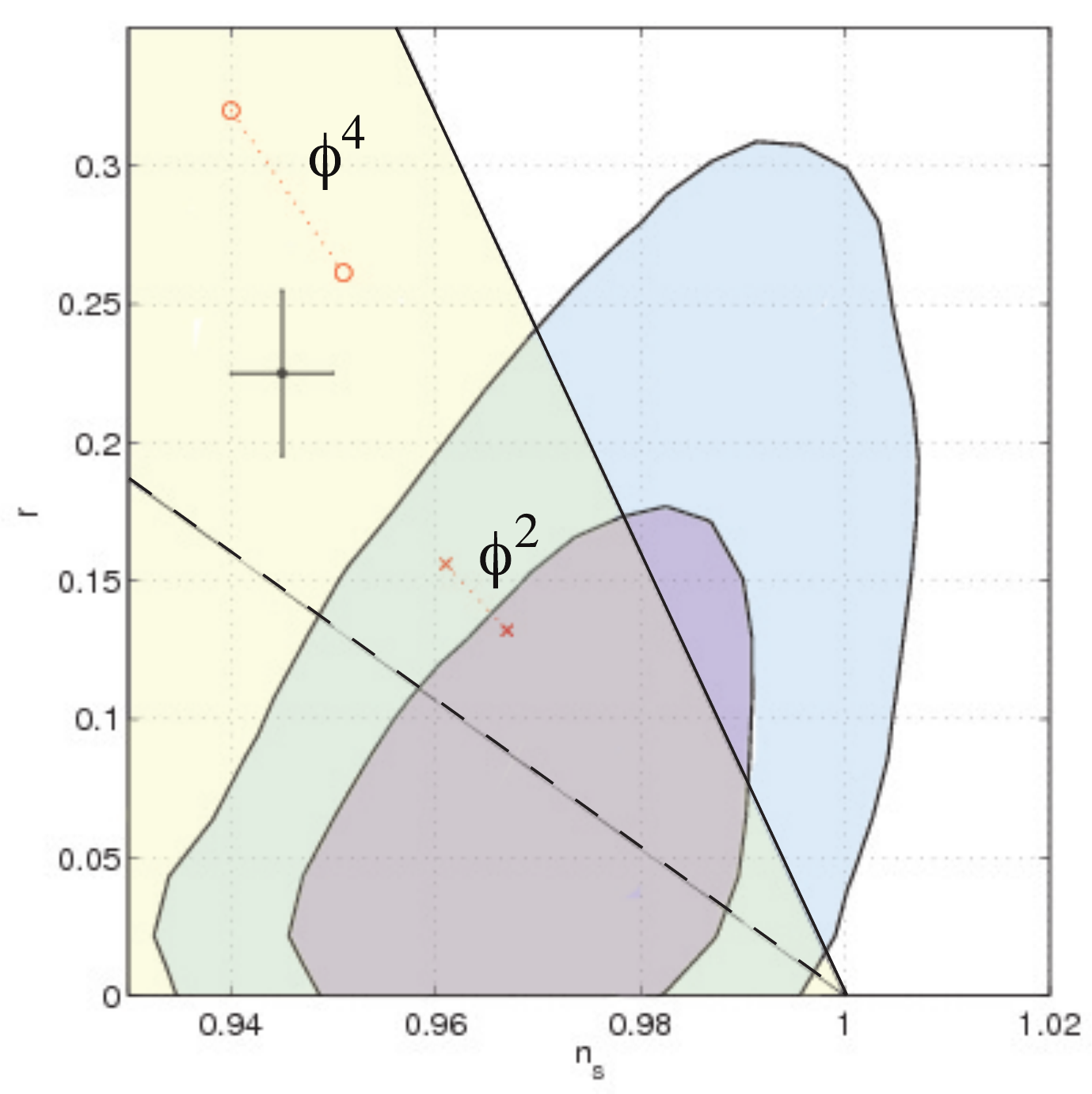}}
\caption{WMAP7+BAO+$H_0$ 68\% and 95\% CL marginalized contours in the $n_s$-$r$ plane
together with estimated Planck error bars.  The (red) line segments show the predictions for $V (\phi) = m^2\phi^2$ and $V (\phi) = \lambda \phi^4$ for the number $N$ of e-folds before the end of inflation at which a mode crossed outside the horizon in the range $N = [50, 60]$.
Only models predicting observables outside the yellow
region (black solid line) can be uniquely classified according to the inflationary zoology when an unresolved curvaton is
present. Small field models lie below the dashed black line.}
\end{figure}
 
\begin{figure*}[h!]
\begin{center}
\subfigure[]{
\includegraphics[scale=0.25]{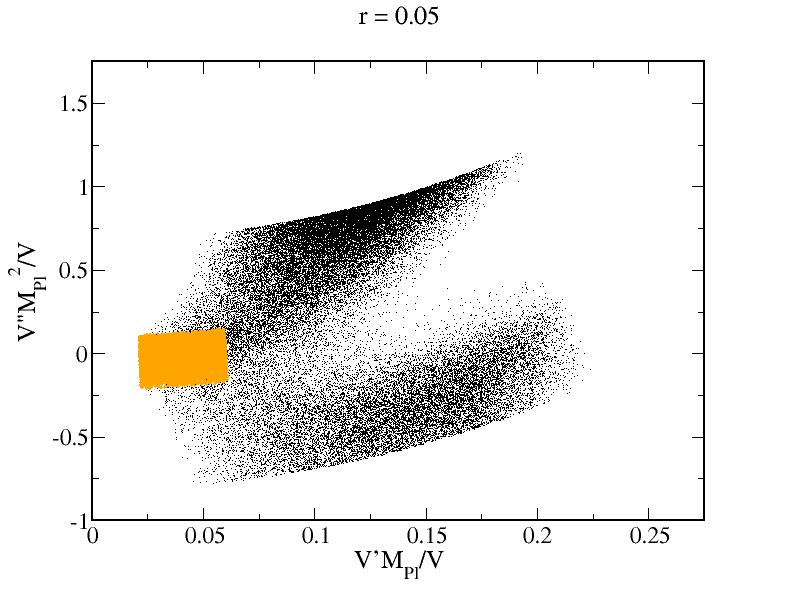}}
\subfigure[]{
\includegraphics[scale=0.25]{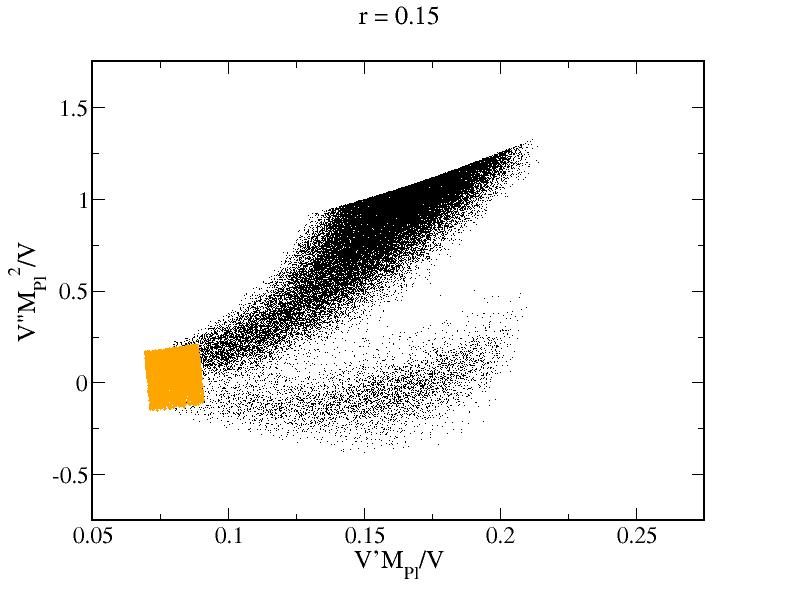}}
\subfigure[]{
\includegraphics[scale=0.25]{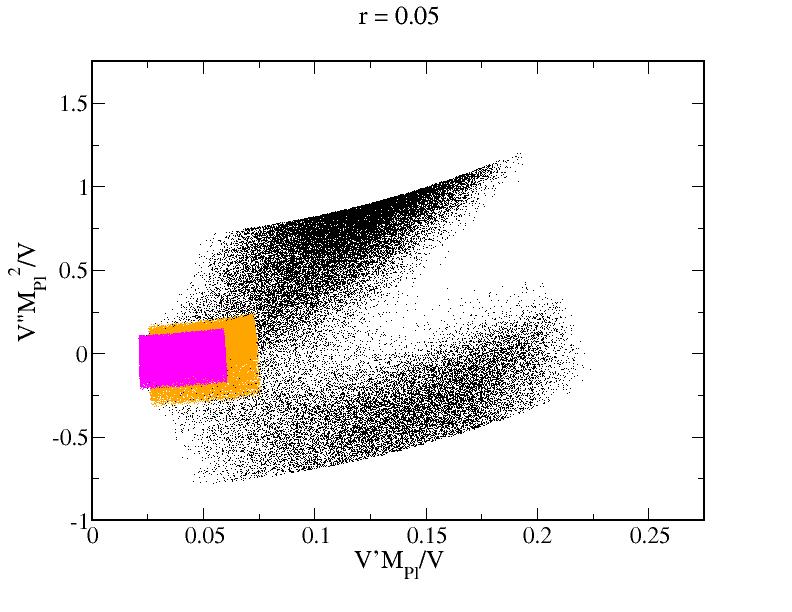}}
\subfigure[]{
\includegraphics[scale=0.25]{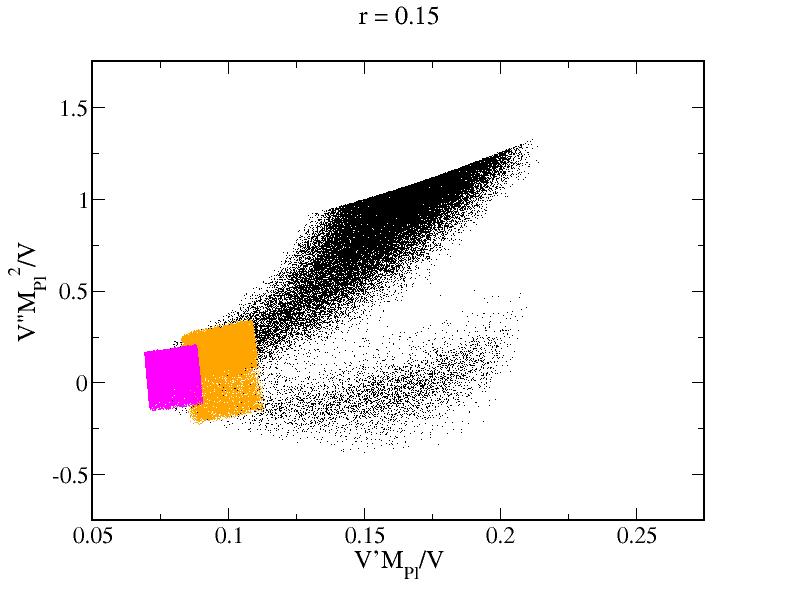}}
\subfigure[]{
\includegraphics[scale=0.25]{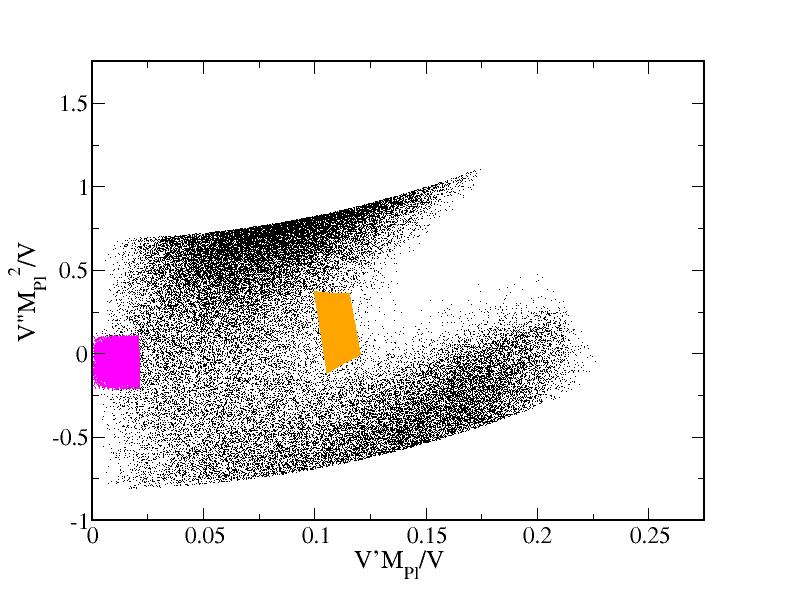}}
\subfigure{
\includegraphics[scale=0.25]{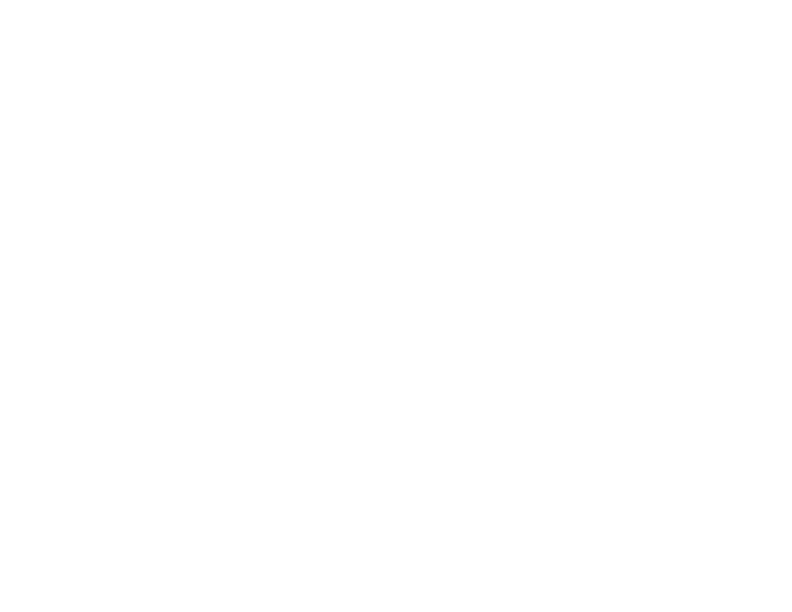}}
\end{center}
\caption{Monte Carlo results (orange) incorporating a detection of CDM isocurvature perturbations: uncorrelated (a) and (b),
arbitrary correlation (c) and (d), anti-correlated (e).  Models from Figure 2 \S \ref{base}, in which an
isocurvature contribution is absent, are included for comparison.}
\end{figure*}

\subsubsection{Detection of CDM Isocurvature Modes}
We now extend the results of the previous section by considering a detection of an isocurvature
contribution to the density perturbation.  We assume the same detections of $r$, $n_s$, and $dn_s/d{\rm ln}k$, and analyze the three cases discussed in \S \ref{sec:iso}: an uncorrelated CDM mode
($\beta = 0$), an
anti-correlated CDM mode ($\beta = -1$), and a CDM mode with an arbitrary correlation angle.  We present results
of the Monte Carlo in Figure 5 for the same fiducial values of $r$ as considered previously.  We plot the new
constraints in orange and we include
the constraints that are obtained in the absence of an isocurvature detection (Figure 2) for
comparison.  It should be clear that the detection of
an isocurvature mode rules out single field inflation and breaks the degeneracy; however, we will
routinely make
use of the constraints obtainable on single field inflation in the absence of any
degeneracy-breaking observables as a benchmark against which to gauge the quality of the
reconstruction. 

The most well-constrained case is that of the uncorrelated CDM mode, Figure 5 (a) and (b). In fact, this case is as
well-constrained as single field inflation; the uncertainty in the amplitude $|\mathcal{S}_{\rm CDM}/\zeta|^2$ in Eq.
(\ref{uncorr}) at Planck precision ($\sim$ 4\% \cite{Bucher:2000hy}) is subdominant relative to the errors on the other power spectrum
observables.  The function $\tilde{f}(\sigma)$ is well-constrained: $\Delta \tilde{f}^2(\sigma) \approx 5$. 

We present the results of the arbitrarily correlated CDM mode in Figure 5 (c) and (d).  We have chosen
$\beta = -0.6$ as a fiducial value with a Planck-precision error $\Delta \beta \sim 0.04$ \cite{Bucher:2000hy}.   
The constraints on $V(\phi)$ remain significantly improved relative to the case in which isocurvature modes are not
detected, but deteriorate slightly relative to a detection of uncorrelated modes, with $\Delta
\tilde{f}^2(\sigma) \sim 250$.

The results of a detection of an anti-correlated mode are presented in Figure 5 (e).  There is only one
observational window to portray, since $r \rightarrow 0$ in the limit that the isocurvature mode becomes
anti-correlated (c.f. Eq. (\ref{rb})).  In this limit, $\tilde{f}^2(\sigma)\epsilon \rightarrow \infty$, and  the potential reduces to the form Eqs.
(\ref{curvlimit}-\ref{curvlimit3}).  Since the function $\tilde{f}(\sigma)$ drops out of the
reconstruction, the errors on $V'/V$ and $V''/V$ are due only to the
uncertainties in the spectral parameters $r$, $n_s$, and $dn_s/d{\rm ln}k$.  The curvaton
reconstruction is consequently comparable in precision to single field inflation in the absence of
an isocurvature detection. 
Lastly, since the detection of an isocurvature component rules out single field inflation, there is no
degeneracy problem, and no need to revisit the zoology classification in this case.


\subsubsection{Detection of Non-Gaussianity}
We next extend the results of \S \ref{base} by considering a detection of non-Gaussianities.  We again assume the
same detections of $r$, $n_s$, and $dn_s/d{\rm ln}k$ as previously; we do not consider a detection of
isocurvature modes in this section.  
Even with the most optimistic limits on the nonlinearity parameters: $\Delta f^{\rm local}_{NL} \sim 1$ (68\% CL)
\cite{Carbone:2010sb,Desjacques:2010nn} and $\Delta \tau^{\rm local}_{NL} \sim 560$ (95\% CL) \cite{Kogo:2006kh},
we find that \it a detection of both the bispectrum and trispectrum does not improve constraints relative to those found in
\S \ref{base} for the case in which non-Gaussianities are not detected\rm. The reason is simply that the projected
uncertainty in $\tau^{\rm local}_{NL}$ is too large to sufficiently constrain $r_\sigma$, with the result that even
a precision
detection of $f^{\rm local}_{NL}$ is incapable of sufficiently constraining
$\tilde{f}^2(\sigma)$, with a resulting
uncertainty
$\Delta \tilde{f}^2(\sigma) \approx 1500$. 
This is an example of a case where the degeneracy is broken
but the inversion problem persists.   
\begin{figure*}[htp]
\label{zoo2}
\centering
$\begin{array}{cc}
\subfigure[]{
\includegraphics[width=0.49 \textwidth,clip]{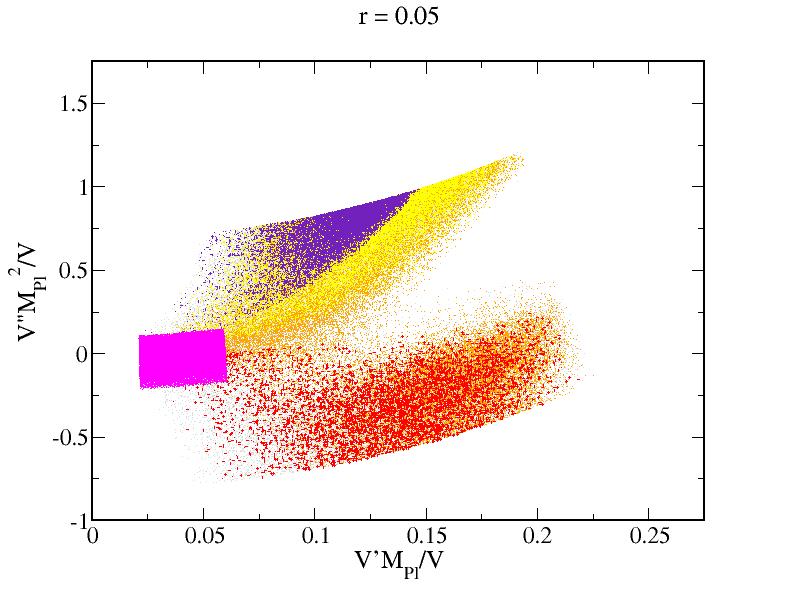}}
\subfigure[]{
\includegraphics[width=0.49 \textwidth,clip]{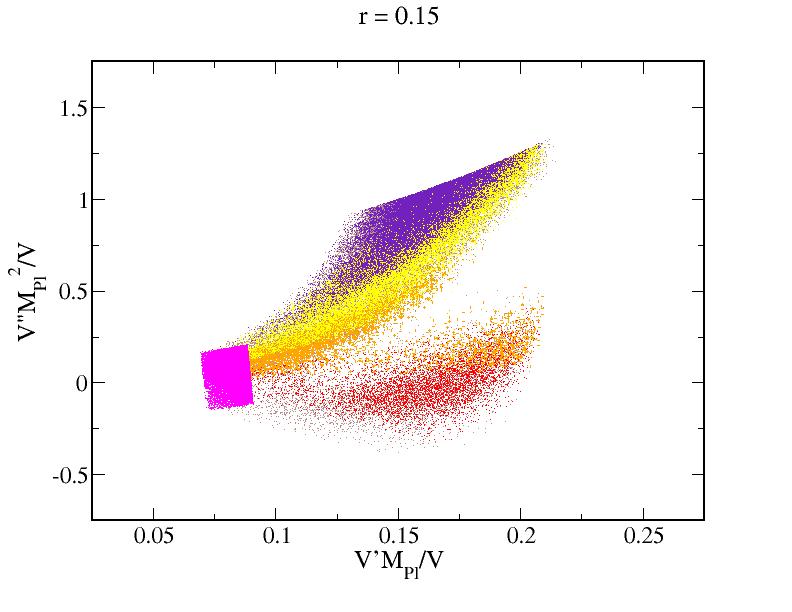}}
\end{array}$
\caption{Monte Carlo results for the case in which a scale dependent non-Gaussianity is detected in the CMB at Planck precision.  For each fiducial value
of $r$ we consider four fiducial dependencies: $n_{NG} = -0.2$ (purple), $-0.1$ (yellow), $0$ (orange), and $0.1$ (red).  From Figures 2 (b) and (c), we include the
degenerate curvaton (grayed-out for clarity) and single field models (magenta) for reference.}
\end{figure*}

While measurements of $f_{NL}$ and $\tau_{NL}$ do not enable a successful inversion, a measurement of the scale dependence of
non-Gaussianities by itself can be utilized in the reconstruction.  Since $\Delta n_{NG} \sim 1/f_{NL}$ (Eq. (\ref{errornng})), the most optimistic outcome
is  when $f^{\rm local}_{NL}$ saturates its current observational bound, $f^{\rm local}_{NL} \approx 70$.  We consider this case with the four fiducial
values: $n_{NG} = -0.2$, $-0.1$, $0$, and $0.1$ at Planck precision ($\Delta n_{NG} = 0.07$).  We present results in Figures 6 (a) and (b) and note the
improvement over the degenerate case Figure 2 (b) and (c) .  For smaller $f^{\rm local}_{NL}$, the uncertainty in $n_{NG}$ grows and this observable ceases
to be of any help when $f^{\rm local}_{NL} < 35$.  A detection with CMBPol will further improve on these results, with a minimum $f^{\rm local}_{NL} > 17$.   
While the constraints on $V(\phi)$ improve in these cases, we find that $\tilde{f}^2(\sigma)$ is not sufficiently constrained to enable a unique model
classification according to the zoology.

\subsubsection{Measurement of $n_T$}
\label{curvddgw}
The last case that we consider is an extension of \S \ref{base} to include a precision measurement of the tensor spectral index, which will impose constraints on $\tilde{f}^2(\sigma)$ through the consistency relation Eq. (\ref{modcons}).
We consider separately Planck and CMBPol detections of $r$, and investigate the improvement in reconstruction that results from a
direct detection of primordial
gravitational waves, as might be possible with future space-based laser interferometers.
The accuracy with which these
probes will determine $n_T$ on direct detection scales $\propto 0.1-1$ Hz is \cite{Seto:2005qy,Kudoh:2005as}
\begin{equation}
\label{errornt1}
\Delta n_T = \frac{6 \times 10^{-18}}{X\Omega_{GW}h^2},
\end{equation} 
where $\Omega_{GW}$ is the gravity wave density, $h$ is the present day Hubble parameter (in units of 100 km s$^{-1}$ Mpc$^{-1}$) and $X$ characterizes
the particular experiment: $X = 0.25$ (BBO-standard), $X=2.5$ (BBO-grand), and $X=100$ (DECIGO).  The quantity
$\Omega_{GW}h^2$ can be related to the primordial tensor spectrum \cite{Smith:2005mm},
\begin{equation}
\Omega_{GW}h^2 = A_{GW}P_h(k),
\end{equation}
where the transfer function, $A_{GW}$, which governs the evolution of the perturbations up through horizon
re-entry, has the numerical value $A_{GW}  = 2.74 \times 10^{-6}$. 
We take the tensor spectrum to be of the form
\begin{equation}
P_h(k) = P_h(k_0)\left(\frac{k}{k_0}\right)^{n_T + \frac{1}{2}\alpha_T{\rm ln}\left(\frac{k}{k_0}\right)},
\end{equation}
where $\alpha_T = dn_T/d{\rm ln}k$ is the tensor index running.  Although expected to be small and unlikely to be constrained in future experiments, the running becomes important
when considering the large disparity between the scales of CMB probes ($k_0 = 0.01\,  {\rm Mpc}^{-1}$) and laser interferometers (here taken to be $k_* = 6.5 \times 10^{14} \,{\rm
Mpc}^{-1}$).
Our intent is to utilize this measurement of $n_T$ on direct detection scales to constrain the value of $n_T$
at the pivot where the consistency relations are valid.  However, one cannot simply extrapolate the error Eq. (\ref{errornt1}) to CMB scales since this
presumes that $n_T(k_0) = n_T(k_*)$, which does not hold in the presence of running. The uncertainty in $n_T$ at the pivot is therefore 
\begin{equation}
\label{errornt}
\Delta n_T = \left\{\left[\frac{6\times 10^{-18}}{XA_{GW}P_h(k_*)}\right]^2 + \left[\frac{1}{2}\alpha_T{\rm ln}\left(\frac{k_*}{k_0}\right)\right]^2\right\}^{1/2},
\end{equation}
where $\alpha_T \simeq 4\epsilon \eta - 8\epsilon^2$.  Since the running is unknown, it acts as an additional source of uncertainty when trying to match $n_T(k_*)$ to $n_T(k_0)$.
The running is not, however, completely free to vary.  It must be such that the tensor amplitude on direct detection scales gives the correct amplitude of
gravitational waves, $\Omega_{GW}$, to within experimental error of the fiducial value.  The 1$\sigma$ error on $\Omega_{GW}$ is approximately set by the
signal-to-noise ratio of the experiment, $\Delta \Omega_{GW} \approx (S/N)^{-1}$, with $S/N = X\Omega_{GW}/10^{-18}$ \cite{Seto:2005qy}.  We take as our
fiducial values $n_T(k_*) = -r/8$ and $\Omega_{GW}h^2 = A_{GW}rP_\Phi(k_0)(k_*/k_0)^{n_T(k_*)}$, consistent with a power
law tensor spectrum. 

As discussed previously, as long as $r$ and $n_T$ are consistent with Eq.
(\ref{modcons}) to within experimental error, the relation can be used in the reconstruction.
For ease of comparison, we choose to base our reconstructions on fiducial values for $r$ and $n_T$ that agree
with the single field relation; other choices will result in similar constraints on
$\tilde{f}^2(\sigma)$, since for a given value of $r$, $\Delta n_T$ depends only weakly
on the actual value of $n_T$ chosen.  
\begin{figure*}[htp]
\label{MC}
\goodgap \goodgap \goodgap \goodgap \goodgap \goodgap \goodgap \goodgap \goodgap                                                                                  
\goodgap \goodgap \goodgap \goodgap \goodgap \goodgap \goodgap \goodgap
\subfigure[]{
\includegraphics[width=0.32 \textwidth,clip]{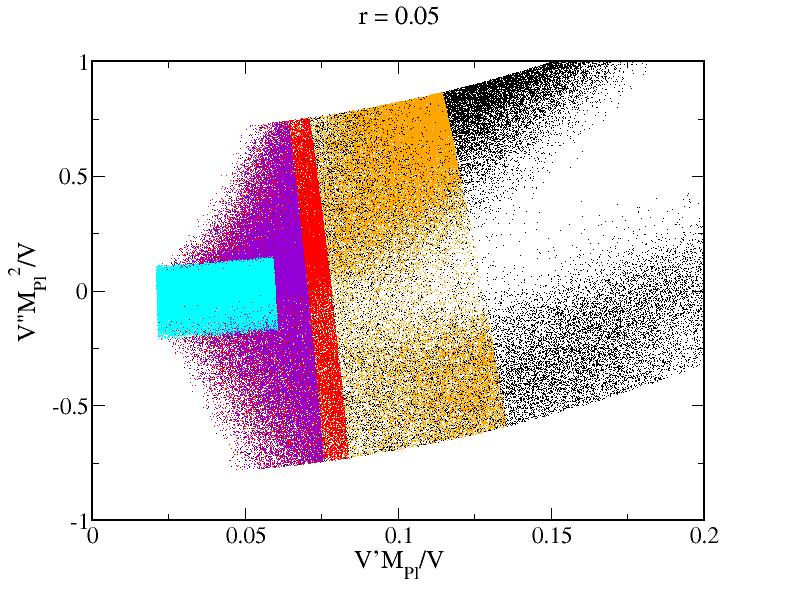}}
\subfigure[]{
\includegraphics[width=0.32 \textwidth,clip]{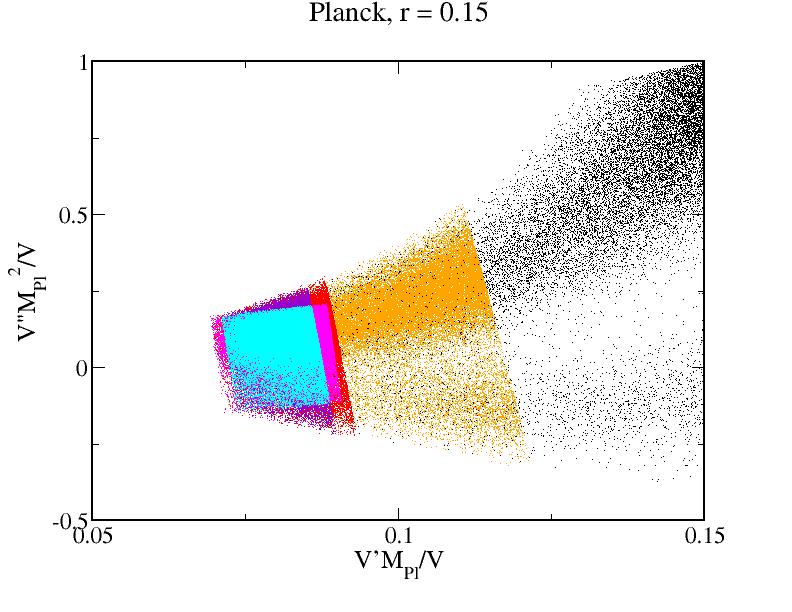}}
$\begin{array}{ccc}
\subfigure[]{
\includegraphics[width=0.32 \textwidth,clip]{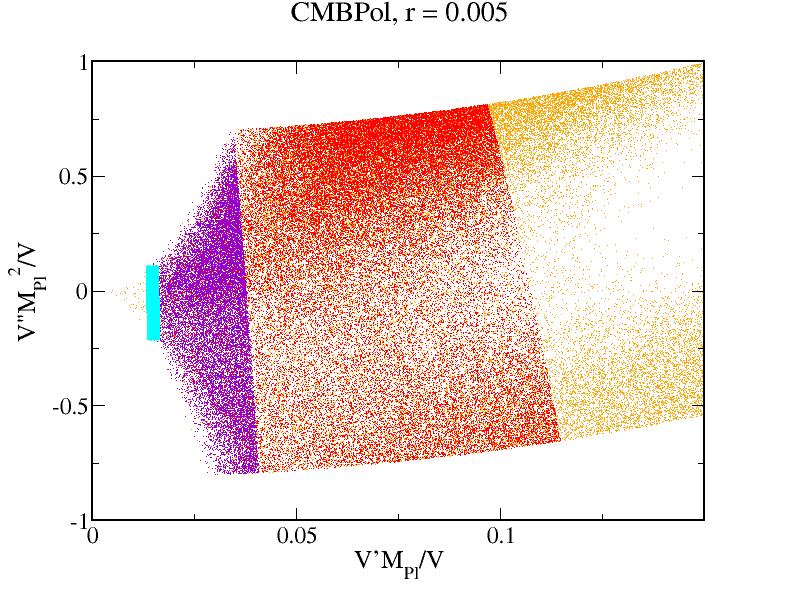}}
\subfigure[]{
\includegraphics[width=0.32 \textwidth,clip]{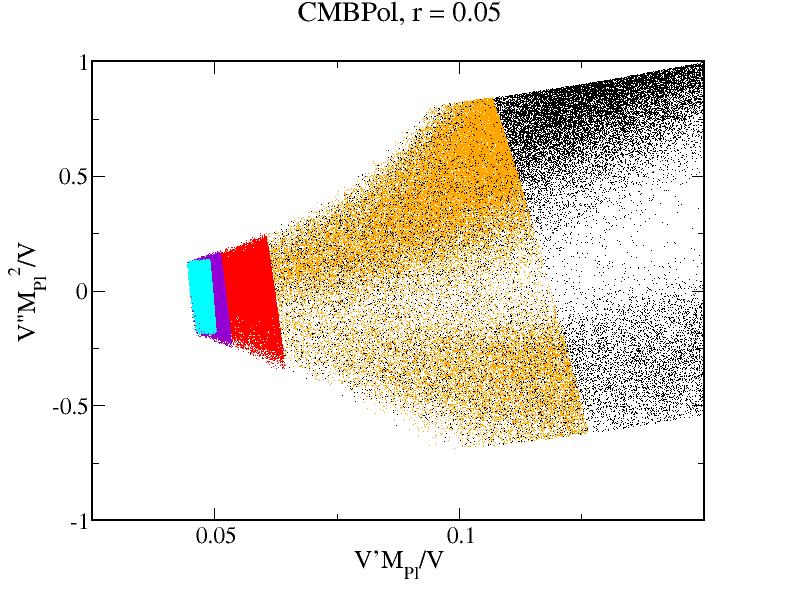}}
\subfigure[]{
\includegraphics[width=0.32 \textwidth,clip]{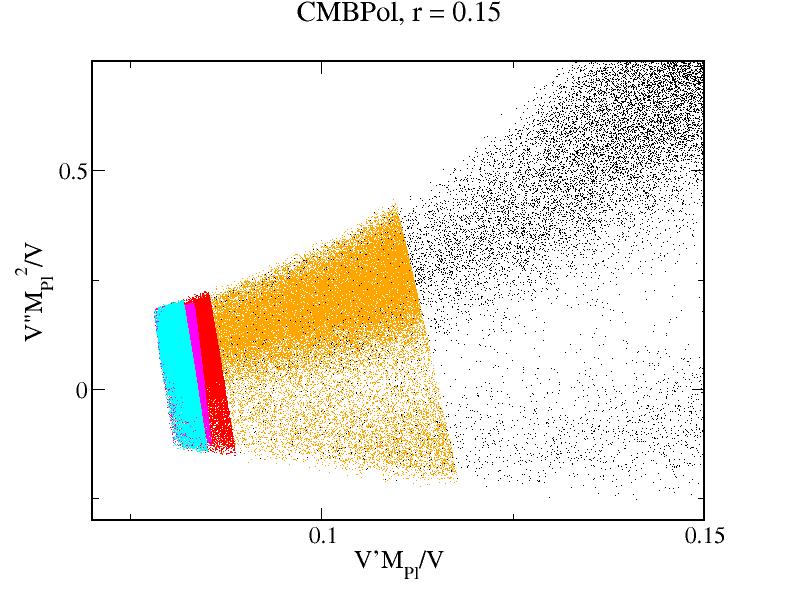}}
\end{array}$
$\begin{array}{ccc}
\subfigure[]{
\includegraphics[width=0.32 \textwidth,clip]{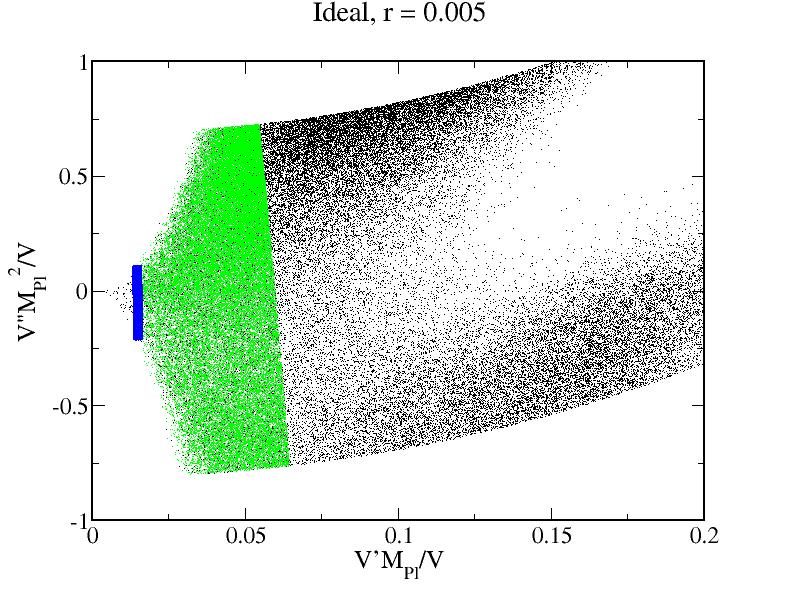}}
\subfigure[]{
\includegraphics[width=0.32 \textwidth,clip]{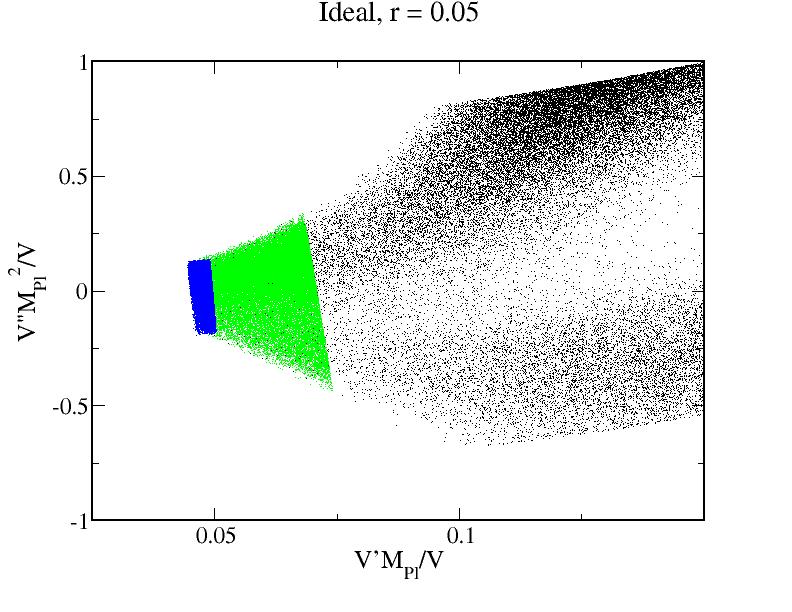}}
\subfigure[]{
\includegraphics[width=0.32 \textwidth,clip]{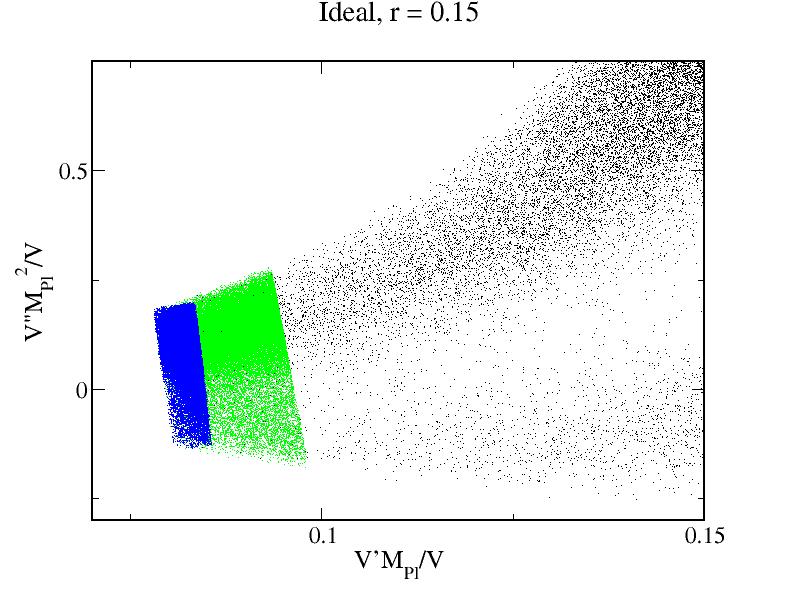}}
\end{array}$
\caption{Monte Carlo results for single field and curvaton models with a direct detection of tensors at three
different precisions: BBO-standard, BBO-grand, and                                                              
DECIGO, for Planck (top row) and CMBPol (middle row).  We present results at three fiducial values: $(r=0.005$, $n_T = -0.00065)$, $(r = 0.05$, $n_T =
-0.00625)$, and $(r = 0.15$, $n_T = -0.01875)$. 
We include the curvaton and single field models without a measurement of $n_T$ and overlay the reconstructions from direct detection at each different precision.
For curvatons: BBO-standard (orange), BBO-grand (red), DECIGO (purple); for singe field we only present results from DECIGO (cyan), since the other
reconstructions do not improve on the case without direct detection (magenta).  In the last row we present projections in which $n_T$ is determined by an
ideal detection of B-modes on CMB scales rather than a direct detection experiment: curvatons (green) and single field (blue).}
\end{figure*}

We present Monte Carlo results for Planck in Figure 7: (a) and (b) correspond to fiducial values ($r = 0.05$, $n_T = -0.00625$) and ($r = 0.15$, $n_T =
-0.01875$), respectively, and we compare the reconstruction projections of a 
direct detection of tensors with BBO-standard, BBO-grand, and DECIGO. We retain the curvaton (black) and single field (magenta) models from Figures 2
(b) and (c) and overlay the reconstructions from each direct detection experiment.  For $r=0.05$, direct detection offers no improvement for single field inflation since $\Delta r < \Delta n_T$.  However, there is improvement
in $V'/V$ for curvatons: BBO-standard (orange), BBO-grand (red), and DECIGO (purple).  Overall, constraints are improved for $r=0.15$, where we find that
single field uncertainties are reduced with DECIGO (cyan).  The improvement for curvatons is similar to that seen in Figure 7 (a), and the curvaton and single field reconstructions are
almost equivalent with DECIGO.  For both curvatons and single
field, we note the rather insignificant improvement of DECIGO over BBO-grand, despite the factor of forty increase in sensitivity offered by DECIGO.  
We consider the projections for CMBPol in Figure 7 (c)-(e).  At $r=0.15$, the errors on the
curvaton reconstruction are reduced to those of single field inflation (purple points behind cyan).  Constraints are
weakest for $r=0.005$ -- BBO-standard
offers no improvement over the degenerate case.

It is important to emphasize that the improvement in reconstruction that results from the direct detection of gravity waves relies on 
optimistic outcomes of missions still in the planning stages, and should therefore be viewed as a best-case scenario.  In the future, the tensor index
might instead be constrained by a more accurate measurement of B-modes on CMB scales.  For completeness, we consider projections of an ideal detection of
B-modes on CMB scales in Figures 7 (f)-(h).  The constraints are competitive with direct detection, achieving an accuracy close to that of
CMBPol+BBO-grand.  The uncertainties in $n_T$ for the ideal experiment are taken to be $\Delta n_T = 0.01$ at $r=0.005$, $\Delta n_T = 0.009$ at $r=0.05$, and $\Delta
n_T = 0.007$ at $r=0.015$ \cite{Zhao:2009mj}.
\begin{figure*}[htp]
\label{Zzoo}
\goodgap \goodgap \goodgap \goodgap \goodgap \goodgap \goodgap \goodgap \goodgap
\goodgap \goodgap \goodgap \goodgap \goodgap \goodgap \goodgap \goodgap \goodgap
\goodgap \goodgap \goodgap \goodgap \goodgap \goodgap \goodgap \goodgap \goodgap
\goodgap \goodgap \goodgap \goodgap \goodgap
\subfigure[]{
\includegraphics[scale=0.20]{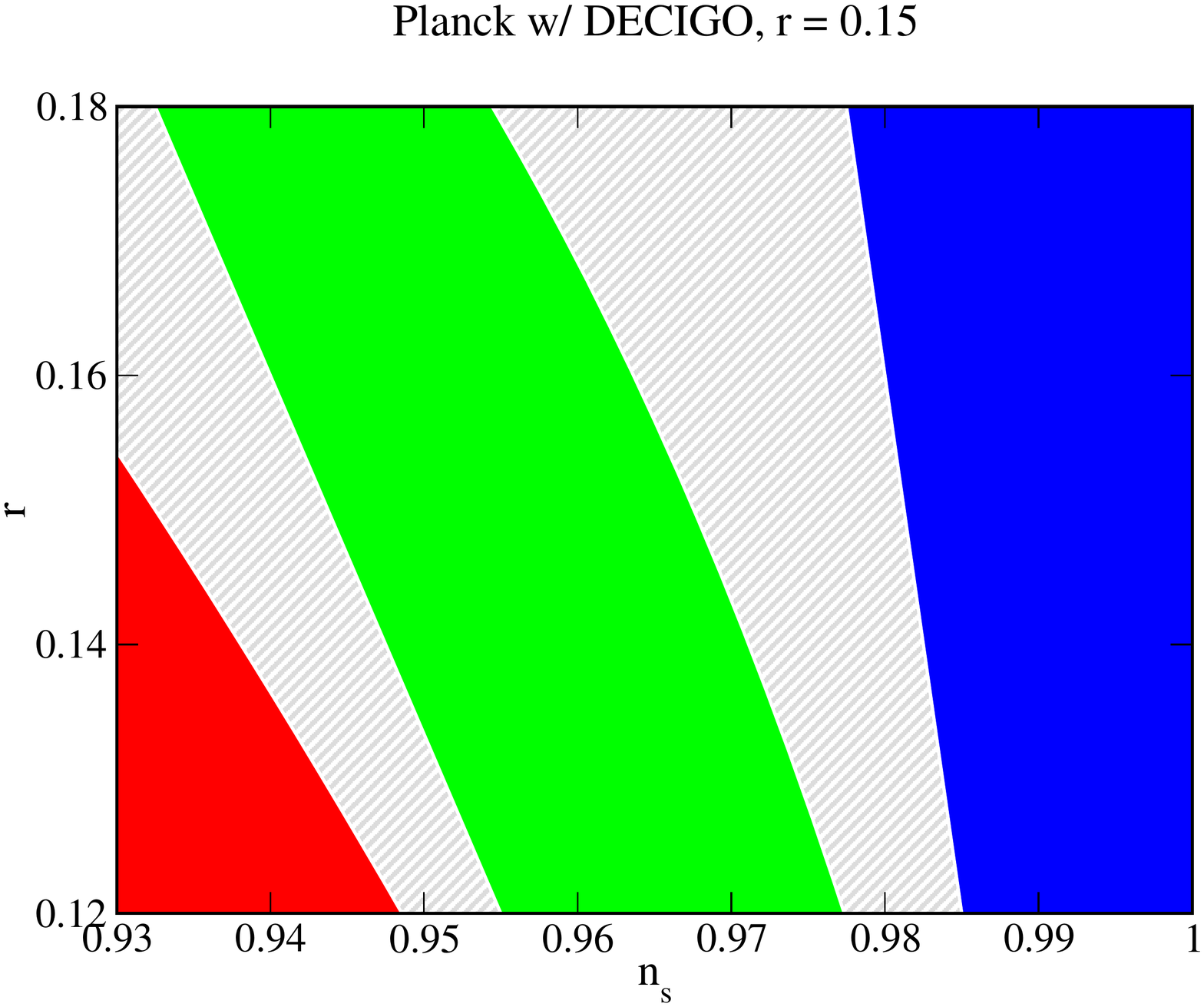}}\\
$\begin{array}{cc}
\goodgap \goodgap \goodgap \goodgap \goodgap \goodgap \goodgap \goodgap \goodgap
\goodgap \goodgap \goodgap \goodgap \goodgap \goodgap \goodgap 
\subfigure[]{
\includegraphics[scale=0.20]{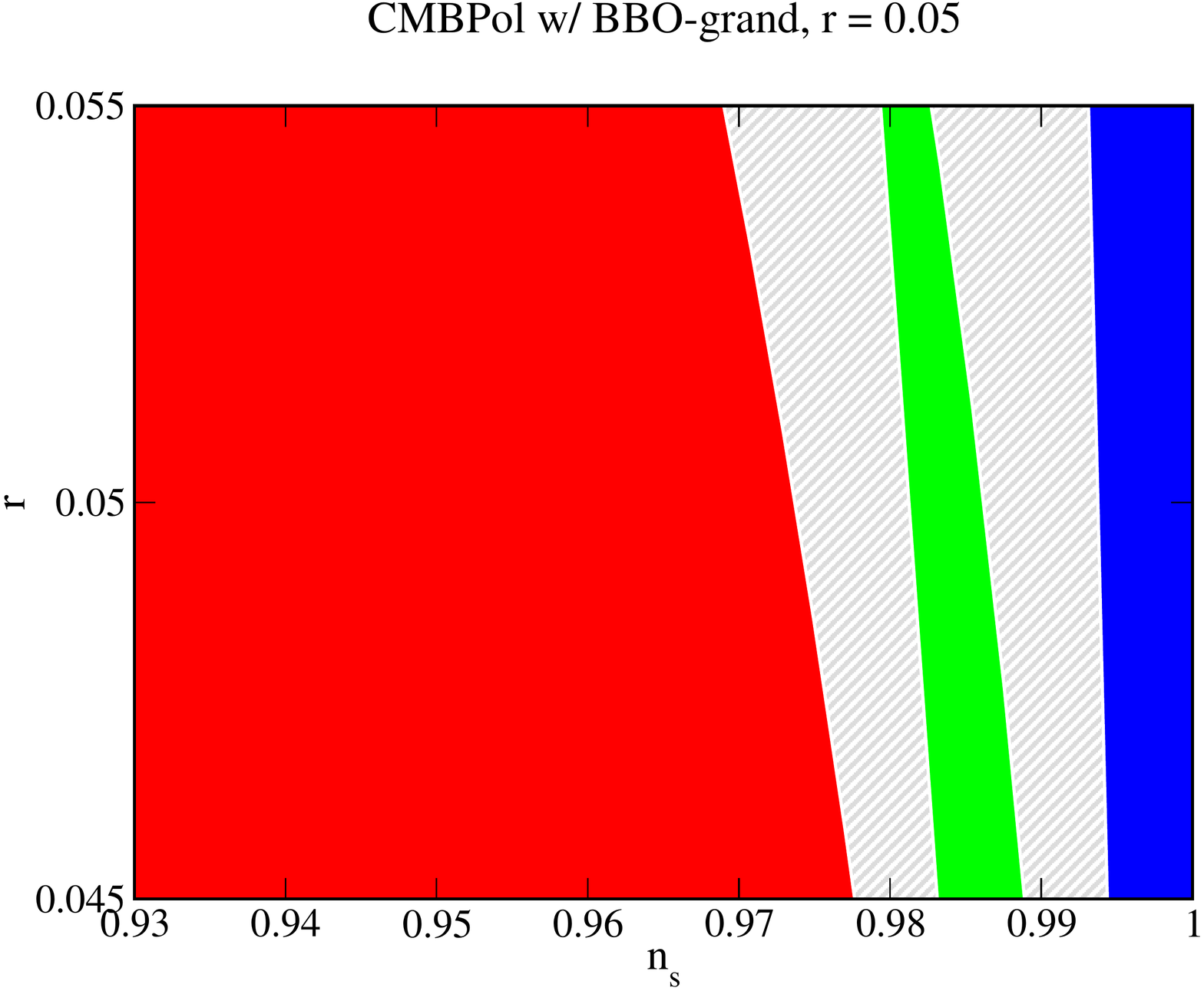}}
\subfigure[]{
\includegraphics[scale=0.20]{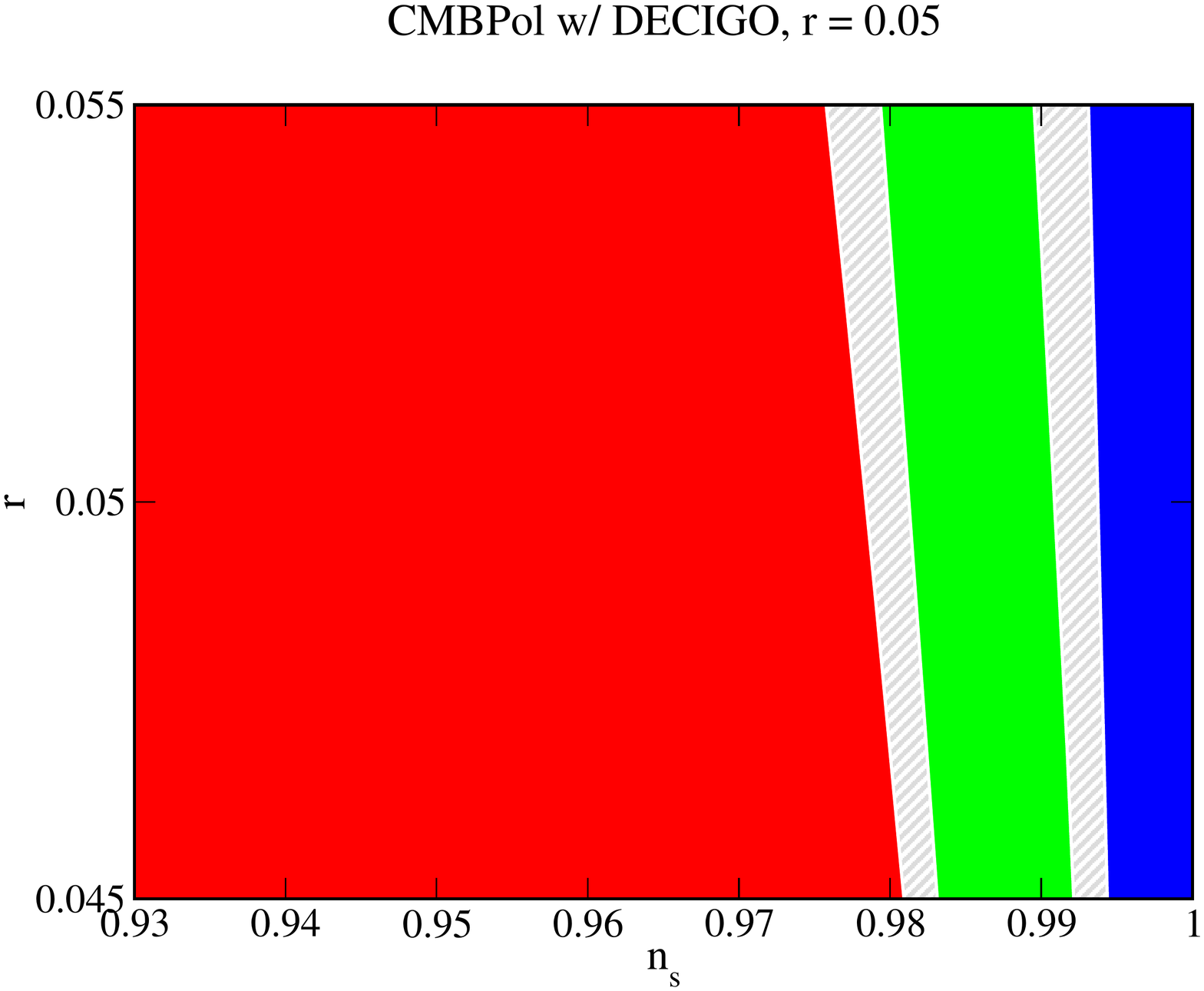}}
\end{array}$
$\begin{array}{ccc}
\subfigure[]{
\includegraphics[scale=0.20]{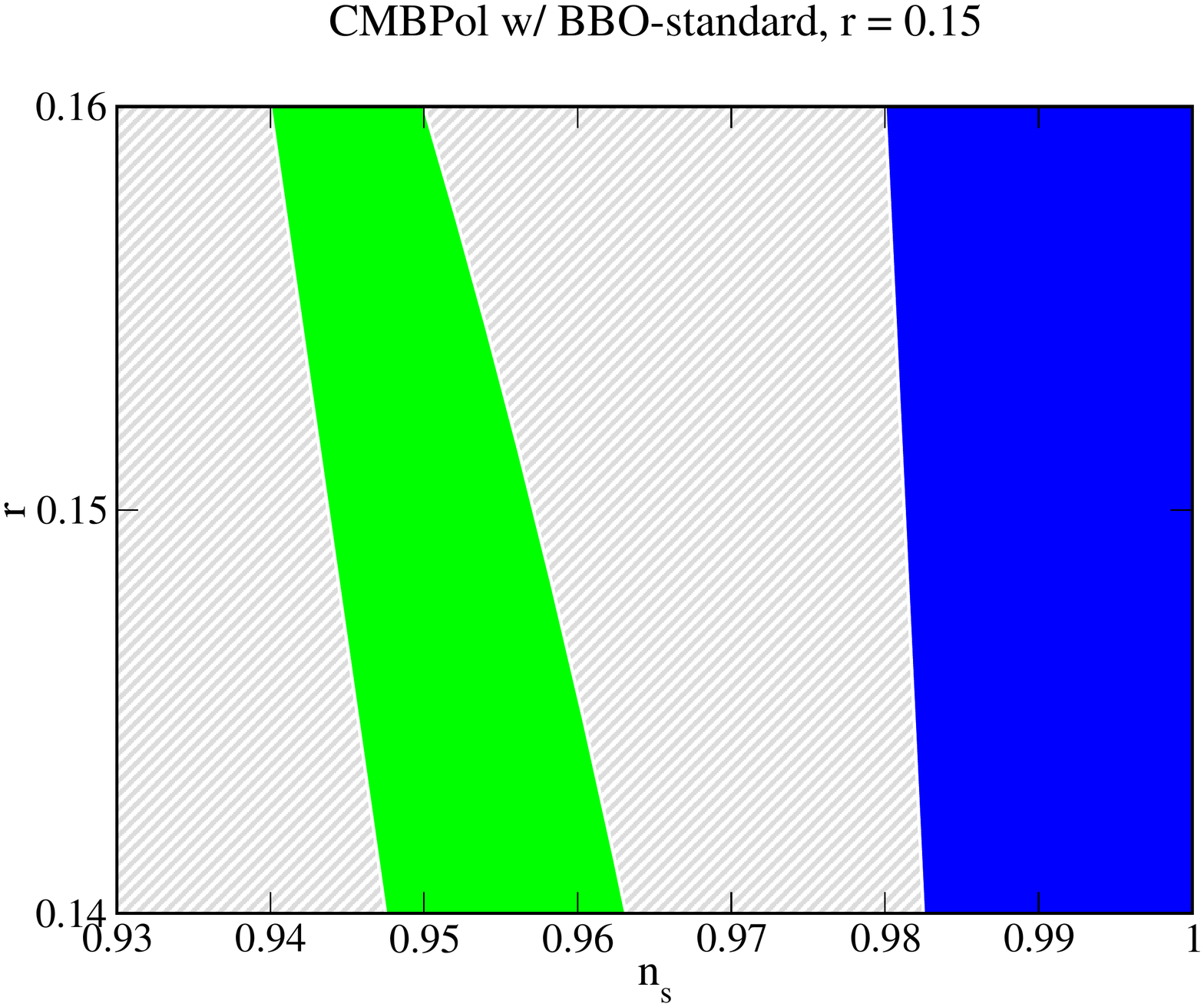}}
\subfigure[]{
\includegraphics[scale=0.20]{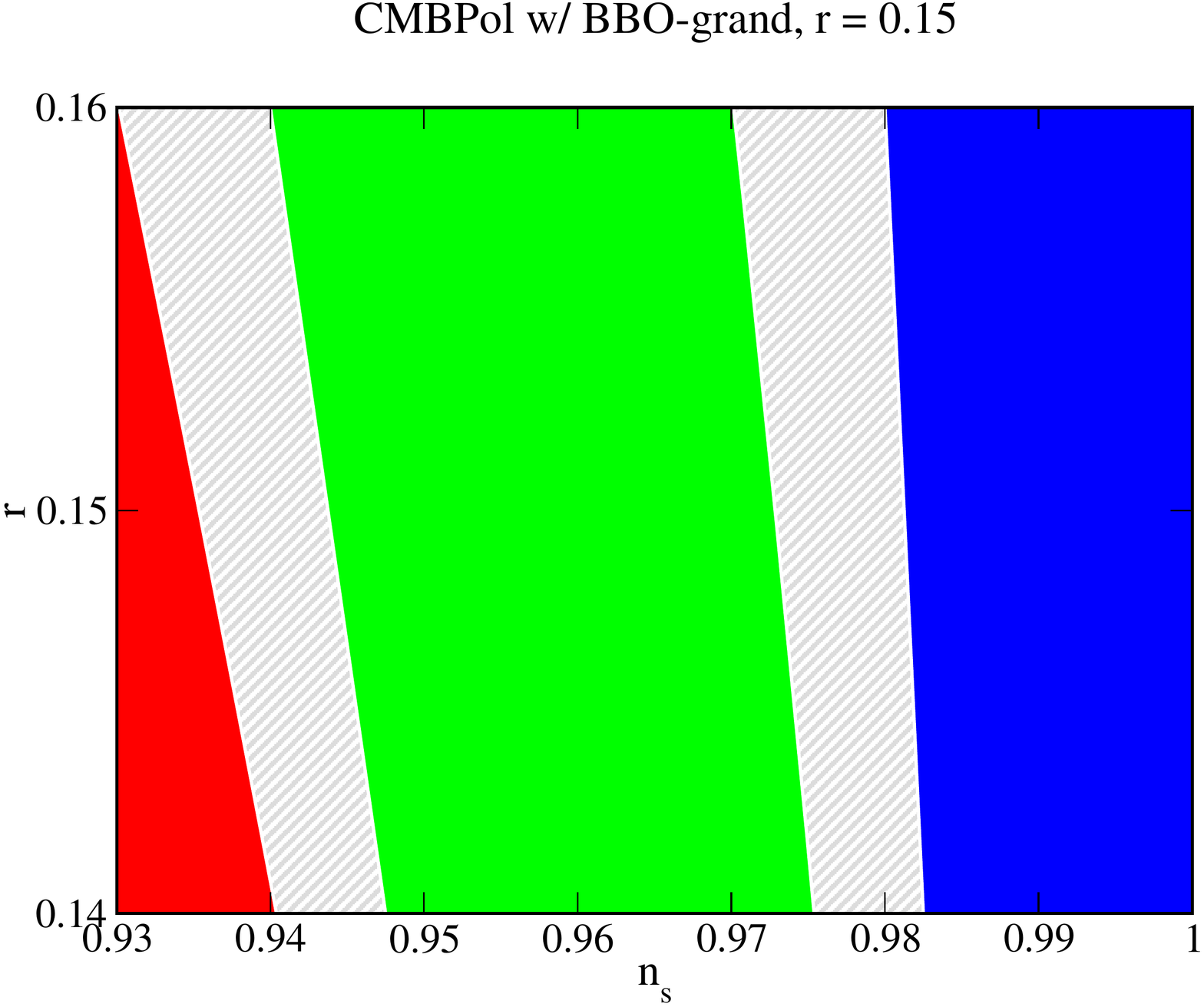}}
\subfigure[]{
\includegraphics[scale=0.20]{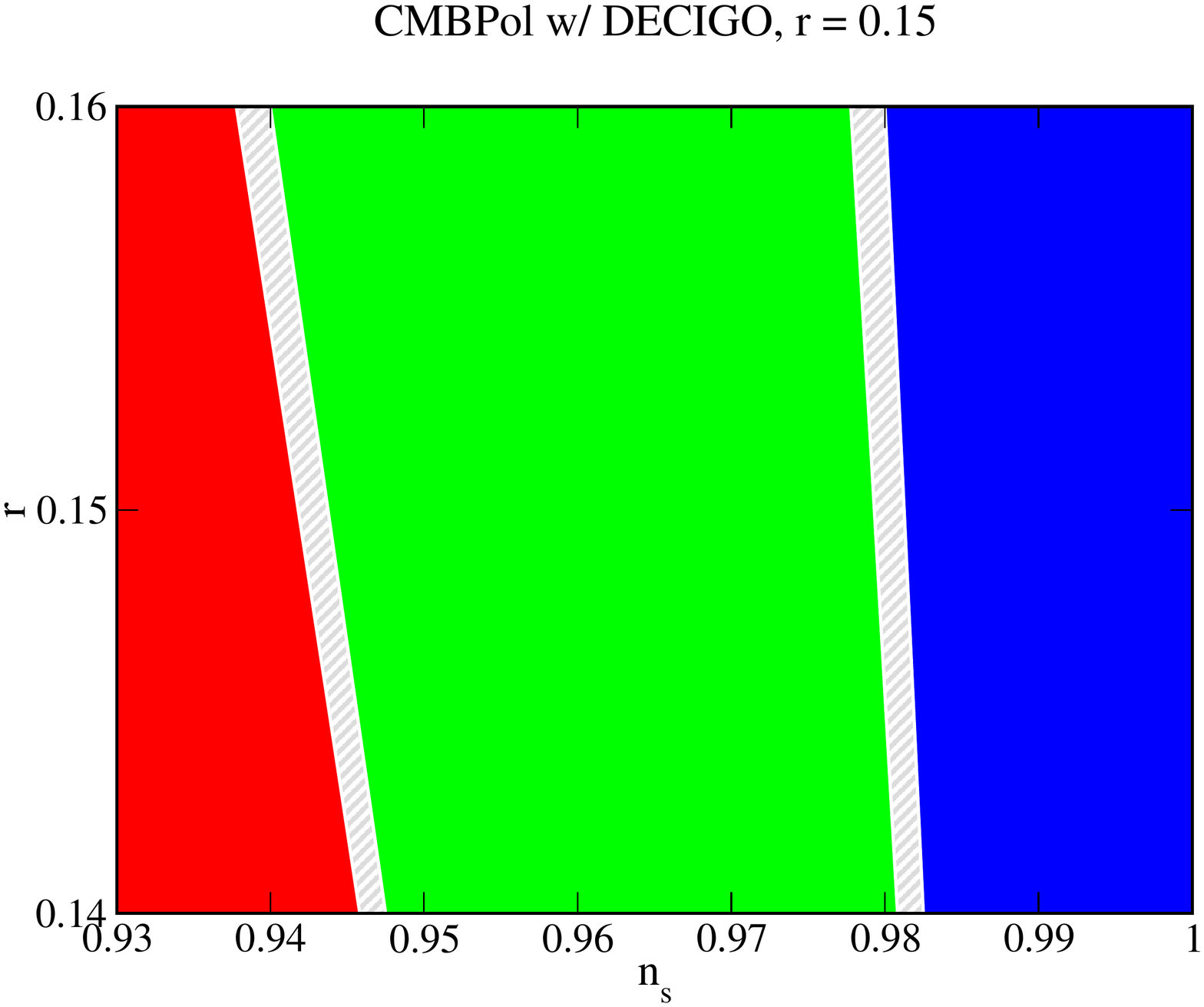}}
\end{array}$
\caption{Zoology in the presence of the curvaton degeneracy for different quality
measurements of $r$ and $n_T$, at $r=0.05$ and $r=0.15$.  Degenerate regions are shown in gray.  Experiments not shown are fully degenerate.}
\end{figure*}


Given the marked improvement in constraints that a direct detection affords, the expectation is that some of the uniqueness of model classification,
characteristic of single field inflation, might be retained even in the presence of curvatons.  The uncertainty in the function
$\tilde{f}^2(\sigma)$ determines an approximate zoology through Eqs. (\ref{line1}) and (\ref{2}), and an accurate measurement of $n_T$ can be used to constrain it through the
consistency relation Eq. (\ref{modcons}).
In Figure 8, we present the zoology for the cases considered in Figure 7.  Regions in which the
degeneracy persists -- where different classes of single field and curvaton models make the same observable predictions -- are shaded in gray.
With Planck, the variation in $\tilde{f}^2(\sigma)$ is large for $r=0.05$ for each $n_T$ detection, and the zoology is degenerate
(Figure 3 (b)) with
hybrid models the only unique class.  
While the degeneracy is reduced with Planck+DECIGO at $r=0.15$ (Figure 8 (a)), there is little room for a unique classification: the 1$\sigma$ error on $r$
spans the entire y-axis, so small field models cannot be uniquely classified, while large field models might be identified for a precise measurement of
$n_s \approx 0.955$.  
With CMBPol, however, the
zoology for $r=0.05$ is improved with BBO-grand and DECIGO (Figure 8 (b) and (c)); with DECIGO, there exist regions in which a unique classification of hybrid, small, and large field
models is possible.  
CMBPol at $r=0.15$ is shown in Figures 8 (d)-(f).
Unsurprisingly, the best combination is CMBPol+DECIGO for which the zoology almost reduces to that of single field inflation.  Lastly, we note that for
$r=0.005$, the zoology is degenerate for all direct detection sensitivities as well as an ideal B-mode detection.

\section{Discussion}
\begin{table*}
\begin{tabular}{|l|cc|cc|c|c|}
\hline
\hline
Observation & $\Delta V'_{\rm curv}/\Delta V'_{\rm single}$ \,\,&$\Delta V''_{\rm curv}/\Delta V''_{\rm single}$ &$\Delta V'_{\rm curv}/\Delta V'_{\rm deg}$\,\,&$\Delta V''_{\rm curv}/\Delta V''_{\rm deg}$ &$\Delta \tilde{f}^2(\sigma)$ & Zoology unique?\\
\hline
No NG/iso/$n_T$ &4/5 &6.5 &1&1&1500/150 & H \\
NG ($f^{\rm local}_{NL},\tau^{\rm local}_{NL}$)&4/5 &6.5&1&1 &1500/150 & H \\
NG ($n_{NG}=-0.2$,\,$f^{\rm local}_{NL}=70$)&3/5 &2&0.75/1&0.5 &125 & H \\
NG ($n_{NG}=0$,\,$f^{\rm local}_{NL}=70$)&4/5 &6.5/2&1&1/0.5 &125 & H \\
iso ($\beta = 0$) &1 &1 &0.25&0.15&5 & All\\
iso ($\beta = -1$) &1 &2 &0.1&0.3& 0 & All\\
iso ($\beta = -0.6$) &1.25 &1.5&0.3&0.3& 250 & All\\
$n_T$ (Planck + BBOs) &2.5 &6.5/2.2&0.5&0.85/0.5& 1500/100 & H \\ 
$n_T$ (Planck + BBOg) &1.5/1 &6/1.25&0.25&0.8/0.3& 1500/60&H \\ 
$n_T$ (Planck + DEC) &1.25/1 &6/1.25&0.2&0.8/0.3& 1500/50& H/LF,H \\
$n_T$ (CMBPol + BBOs) &50/14/7 &7/6/1.5&1/0.5/0.3&1/0.9/0.5& 1500/300/60 & H/H/LF,H\\ 
$n_T$ (CMBPol + BBOg) &23/4/1.5 &6/2/1&0.5/0.15/0.15&0.9/0.28/0.3 & 1500/175/20 & H/SF,H/LF,H\\ 
$n_T$ (CMBPol + DEC) &7/1.5/1 &5/1.2/1&0.1/0.035/0.05&0.8/0.14/0.3 & 1500/100/10 & H/All/All\\
$n_T$ (ideal) &17/5/3 &6/3/1.5 &0.25/0.2/0.2 &0.7/0.4/0.4 &1500/200/35& H/SF,H/LF,H \\
\hline
\end{tabular}
\caption{Reconstruction results of each case considered in this analysis.  In the first
set of columns the errors on the potential coefficients in the presence of curvatons are given relative to those expected
from single field inflation; in cases where the observation rules out single field inflation, the values are given for
reference only.  In the second set of columns these errors are given relative to the
worst-case degeneracy, {\it i.e.} no detections of non-Gaussianity, isocurvature modes, or
$n_T$.  When constraints depend on the fiducial value of $r$ chosen, we provide results separated by a
slash, with the constraints for $r=0.005$ followed by those for $r=0.05$ and $r=0.15$ (for Planck only these latter two values are relevant);  if only one number is
provided then there is no difference.  In the last column, `All' indicates that all three zoology classes
can be uniquely reconstructed; otherwise, only those classes that can be uniquely reconstructed are listed.  NG: non-Gaussianities, iso: isocurvature, H:
hybrid, LF: large field, SF: small field.}
\end{table*}
We have investigated the degeneracy problem that arises when the primordial power spectrum is generated by something other than the inflaton.   
We analyzed the curvaton as an example of this class of models.  The curvaton was
selected for our case study because it has been analyzed extensively in the literature and is phenomenologically similar to other scenarios in this class, such as modulated reheating. 
While the curvaton is representative of these models, we have analyzed but one corner of the potentially degenerate model space, and our results should
therefore be considered a best-case scenario for reconstruction in the absence of any degeneracy-breaking observable. 
The importance of further input from theoretical model building to help diminish the degeneracy and inversion problems should not be underestimated: the more information we have about the underlying fundamental physics behind inflation, the better off we are in terms of the reconstruction program. For example,
if the inflaton has DBI Lagrangian and is non-minimally coupled to gravity, interesting observational signatures can arise \cite{Easson:2009kk}. Under certain circumstances the degeneracy 
between minimally and non-minimally coupled DBI inflationary models can be resolved \cite{Easson:2009wc}. We consider another major source of possible degeneracies that can arise from non-canonical inflation in \cite{debp10b}. 

Here, we first determined the size of the degeneracy by performing Monte Carlo reconstructions on both single field and curvaton models in the absence of observables
that might serve to distinguish the curvaton, such as local non-Gaussianities and/or isocurvature modes.  We chose
$P_\Phi(k_0)$, $r$, $n_s$, and $dn_s/d{\rm ln}k$ as base observables common to both models and found that reconstruction is degraded in the presence of an
unresolved curvaton because it precludes a unique inversion of the observables to obtain the potential.  
We then introduced additional observables into our study: isocurvature modes, local non-Gaussianities, and the tensor spectral index.  While the former two
observations break the degeneracy by ruling out single field inflation and lending support to the curvaton, our main focus was to study how well a set of
degeneracy-breaking observables can be inverted to reconstruct the scalar potential.  

A comprehensive view of our results is presented in Table 1.  We list
the size of the errors on $V'/V$ and $V''/V$ that result from a detection of
each observable considered in the analysis.  The errors are given relative to the errors
expected from single field inflation (in the event that the observation rules out single field inflation, constraints are compared to single field as a
benchmark), and in the second set of columns we display the uncertainties from each observation relative to the               
totally degenerate case.  The uncertainty in $\tilde{f}^2(\sigma)$ is most directly relevant to the zoology portraits shown throughout the paper.   
Our main conclusions are as follows:
\begin{itemize}
\item \sl Null-detection. \rm In the absence of a detection of non-Gaussianities or isocurvature modes -- the degenerate case -- constraints on $V'/V$ and
$V''/V$ are each degraded by a factor of 4 and 6.5, respectively, relative to single field inflation.  In terms of the zoology, we find that only models
with observables satisfying $r \geq -8(n_s-1)$ can be uniquely classified.  These are hybrid models (see Figure~4).  Even the improved measurements of $r$ and $n_s$
possible with CMBPol offer scant improvement.

\item \sl Non-Gaussianity. \rm The amplitude of non-Gaussianities, $f^{\rm local}_{NL}$, serves as a useful way to discriminate between the two theories, but the estimator is otherwise
unhelpful for reconstruction.  Even with single field inflation ruled out in this case, a detection of more than the amplitude of the bispectrum is needed for inversion.  We
included the possibility of a measurement of the trispectrum but found that the projected uncertainties of future missions are too large to improve constraints
on $V(\phi)$ beyond the degenerate case.  However, a measurement of the scale dependence of non-Gaussianities offers modest improvement with Planck when
$f^{\rm local}_{NL} > 35$ ($f^{\rm local}_{NL} > 17$ with CMBPol.)   

\item \sl Isocurvature. \rm A detection of an isocurvature contribution to the primordial density perturbation also serves to
break the degeneracy and the amplitude and correlation angle of the mode facilitate a successful reconstruction.  For a CDM isocurvature component, an
uncorrelated mode offers the best constraints.  The zoology becomes virtually identical to that of single field inflation.

\item \sl Gravitational waves. \rm A precise measurement of the tensor index, $n_T$, enables the use of the consistency relations of the degenerate models.   
Even when the values of $r$ and $n_T$ satisfy both the single field and curvaton relations and the degeneracy persists,
a precision measurement of $n_T$ can substantially reduce the degeneracy.  The detection must be of sufficient quality: $\Delta n_T \lesssim
\mathcal{O}(10^{-2})$, with $r \gtrsim 0.01$ offering the best reconstructions; this precision might be achieved with a future direct detection of primordial gravitational waves.  Since the curvaton reconstruction
makes use of the consistency relation, a quality measurement of $r$ is also advantageous.  With a Planck detection of $r$, the degeneracy begins to clear 
for fiducial $r = 0.15$ with DECIGO, but the degeneracy remains for smaller $r$.  However, with a CMBPol detection of $r$, the 
degeneracy resolves partially for $r \gtrsim 0.01$ with BBO-grand and almost completely with DECIGO.  In the case of small tensor/scalar ratio, $r \lesssim
0.01$, we find that with CMBPol+BBO-grand and DECIGO, constraints on $V(\phi)$ improve over the case of no $n_T$ but the zoology remains degenerate.
Lastly, an ideal detection of B-modes on CMB scales offers a reconstruction capability similar to that of
CMBPol+BBO-grand.
\end{itemize}

In summary, a detection of an isocurvature component, the scale dependence of local non-Gaussianities, or a precision
measurement of the tensor spectral index enable successful reconstructions.  
The consistency relation is most powerful when the constraints on $r$ and $n_T$ are optimized -- with CMBPol and DECIGO, respectively.  While the
reconstruction in this case is practically equivalent to single field inflation, we require a favorable observational outcome ($r \gtrsim 0.01$),  and optimistic
foregrounds.  We end by considering the possibility that $r \lesssim 0.01$ and goes unmeasured by Planck.   In the degenerate case, a CMBPol detection of $r \lesssim 0.01$ will constrain the energy scale of inflation, but it will not, by itself,
improve constraints on $V'/V$ or $V''/V$.  In this case, the prospects of experimentally probing the detailed physics of the inflationary mechanism are rather dire.  While
detection of $r$ will provide enthralling evidence for the existence of gravitational waves, we have discovered that in order for CMBPol to adequately reconstruct the shape of $V(\phi)$, it must be paired with a
precision direct detection of primordial gravitational waves.

The resolution of the degeneracy and inversion problems will require a herculean effort on the part of precision observational cosmology: we
will need better probes of B-mode polarization, advanced direct probes of  primordial gravitational waves, and an improved understanding of foregrounds.  It is important that
we understand what we can expect to constrain with these future missions. In this analysis we have taken a small step towards attaining this understanding.

\acknowledgments
It is a pleasure to thank C.~Byrnes, G.~Efstathiou, K.~Enqvist, D.~Holz, W.~Kinney, T.~Kobayashi, E.~Komatsu, L.~M.~Krauss, A.~Liddle, A.~Linde, P.~Lubin, 
V.~Mukhanov, S.~Mukohyama, B.~A.~Ovrut, M.~Salem, M.~Sasaki, B.~ Schlaer, L.~Senatore, E.~Silverstein, F.~Takahashi and N.~Yoshida for helpful discussions. 
The work of DAE is supported in part by the World Premier International Research 
Center Initiative (WPI Initiative), MEXT, Japan and by a Grant-in-Aid for Scientific Research 
(21740167) from the Japan Society for Promotion of Science (JSPS), and by funds from the  Arizona State University Foundation,
and by the National Science Foundation under Grant No. PHY05-51164.

\end{document}